\documentclass[aps,prb,twocolumn,groupaddress,longbibliography,preprintnumbers]{revtex4-2} 

\usepackage{amsmath}
\usepackage{amsfonts}
\usepackage{amssymb}
\usepackage{graphicx}
\usepackage{color}
\usepackage{accents}
\usepackage[colorlinks=true, citecolor=cyan]{hyperref}
\usepackage{enumitem}

\newcommand{\ket}[1]{|#1\rangle}

\makeatletter 
\newsavebox{\@brx}
\newcommand{\llangle}[1][]{\savebox{\@brx}{\(\m@th{#1\langle}\)}%
  \mathopen{\copy\@brx\kern-0.5\wd\@brx\usebox{\@brx}}}
\newcommand{\rrangle}[1][]{\savebox{\@brx}{\(\m@th{#1\rangle}\)}%
  \mathclose{\copy\@brx\kern-0.5\wd\@brx\usebox{\@brx}}}
\makeatother

\newlength{\dhatheight} 

\newtheorem{result}{Result}

\newtheorem{definition}{Definition}

\newcommand{\qed}{\nobreak \ifvmode \relax \else
      \ifdim\lastskip<1.5em \hskip-\lastskip
      \hskip1.5em plus0em minus0.5em \fi \nobreak
      \vrule height0.75em width0.5em depth0.25em\fi}

\usepackage{booktabs}

\begin{document}

\title{Parent Hamiltonians for stabilizer quantum many-body scars}
\author{Shane Dooley} \email[Email address: ]{dooleysh@gmail.com}
\affiliation{School of Theoretical Physics, Dublin Institute for Advanced Studies, 10 Burlington Road, Dublin 4, Ireland.}

\preprint{DIAS-STP-26-01}
\date{\today}

\begin{abstract}
  Quantum many-body scars (QMBS) have attracted considerable interest due to their role in weak ergodicity breaking in many-body systems. We present a general construction that embeds stabilizer states as QMBS of local Hamiltonians. The method relies on a notion of factorizability of Pauli strings on a lattice, which is used to convert stabilizer elements into local, few-body operators that annihilate the stabilizer state. This enables the systematic construction of parent Hamiltonians with zero-energy stabilizer QMBS typically near the middle of the spectrum. The method reproduces several known results in a unified framework, including recent examples of volume-law entangled QMBS, such as the ``rainbow'' QMBS and the entangled antipodal Bell pair state. We also apply the framework to construct examples of stabilizer QMBS with a more complex entanglement structure, such as the cluster state, the toric code state, and a volume-law entangled state we dub the antipodal toric code (ATC) state. Exact diagonalization confirms our results and reveal the stabilizer states as exact eigenstates of their parent Hamiltonian.
\end{abstract}
 

\maketitle

\emph{Introduction.} Many-body Hamiltonians with generic local interactions have energy eigenstates that are ``thermal'' \cite{DAl-16,Mor-18}. Roughly speaking, this means that near the middle of the spectrum the eigenstates are qualitatively similar to random states, e.g., in terms of expectation values of local observables \cite{Deu-91,Sre-94}, their entanglement entropy \cite{Gar-18a}, or their nonstabilizerness (also known as magic) \cite{Tur-25b}. Quantum many-body scars (QMBS) have attracted significant attention as anomalous \emph{nonthermal} eigenstates in an otherwise thermal spectrum \cite{Ser-21a,Mou-22a,Cha-23b}. Such eigenstates can inhibit thermalisation, which has been observed experimentally \cite{Ber-17,Blu-21,Che-22b,Zha-23a,Su-23a,Zha-25a} and may have practical relevance for quantum information processing \cite{Doo-21a,Doo-23a,Doo-25a}. While early examples of QMBS were identified in constrained models such as the PXP chain \cite{Tur-18a,Tur-18b,Lin-19,Cho-19} and projector embedded Hamiltonians \cite{Shi-17a,Mou-18a}, subsequent work has revealed a growing zoo of models with QMBS \cite{Ok-19, Mou-20a, ODe-20, Doo-20b, Got-23a, Doo-22a, Mou-22c, Log-24a, Har-25a}.  

In parallel, stabilizer states occupy a central place in quantum information theory as efficiently characterizable quantum states that can be generated with Clifford circuits \cite{Nie-00}. These include Bell states, Greenberger–Horne–Zeilinger (GHZ) states \cite{Gre-90a}, cluster states \cite{Bri-01a}, the toric code states \cite{Kit-03b}, and other error correcting code states \cite{Ter-15a}. Although stabilizer states can be highly entangled, they are nonthermal since, by definition, they have zero magic in contrast to thermal states, which have magic that scales with the system size \cite{Tur-25b}.

It is well known that any stabilizer state is an eigenstate of its ``stabilizer Hamiltonian'', constructed as a linear combination of stabilizer group elements. However, such stabilizer Hamiltonians often involve multi-qubit interactions or nonlocal interactions that are difficult to realise in the laboratory \cite{Har-25a}. Whether the same stabilizer states can appear as eigenstates of more physically plausible Hamiltonians with few-body local interactions remains an interesting open question. 

In this work, we introduce a general method for embedding stabilizer states as QMBS of Hamiltonians with few-body, local interactions. Our construction is based on a notion of $(k,\ell,b)$-factorizability of Pauli strings. Whenever a stabilizer element admits a $(2,\ell,b)$-factorization, it yields a corresponding $\ell$-local, $b$-body operator that annihilates the stabilizer state. A linear combination all such contributions produces an $\ell$-local, $b$-body parent Hamiltonian whose stabilizer state lies at zero energy, which is typically near the middle of the spectrum. This mechanism thereby provides a systematic route to construct Hamiltonian models with exact stabilizer QMBS.

Our approach can be used to understand several known results in a unified framework, including recent examples of volume-law entangled QMBS \cite{Lan-22a,Chi-24a,Iva-25a}. Going beyond known results, we construct a number of models with stabilizer QMBS, including the cluster state, toric code states, and models with volume-law entangled stabilizer QMBS. In these examples the corresponding parent Hamiltonian is typically local and 2-body, even when the underlying stabilizers involve 3-body operators (for the cluster state) or 4-body operators (for the toric code). 

\emph{Pauli strings on a lattice.} We consider a system of $N$ qubits arranged on a $D$-dimensional lattice. A \emph{Pauli string} is an $N$-qubit operator of the form $\hat{P} = \theta \bigotimes_{n=0}^{N-1} \hat{\sigma}_n^{\mu_n}$ where $\theta \in \{\pm 1 \}$ and $\hat{\sigma}_n^{\mu_n}$ are the single-qubit Pauli operators acting on qubit $n$, with $\mu_n \in \{ I, X, Y, Z \}$. A Pauli string is \emph{$b$-body} if it acts non-trivially on at most $b$ qubits, while a linear combination of Pauli strings (e.g. a Hamiltonian) is called $b$-body if each term is $b$-body. We also say that a Pauli string is \emph{$\ell$-local} if all qubits on which it acts non-trivially are contained in a contiguous subset of at most $\ell$ neighbouring qubits on the lattice. Any linear combination of $\ell$-local Pauli strings is also $\ell$-local.

\emph{Stabilizer states.} A pure state $\ket{\Psi}$ is \emph{stabilized} by the Pauli string $\hat{P}$ if $\hat{P} \ket{\Psi} = \ket{\Psi}$. If $\ket{\Psi}$ is stabilized by $N$ independent and commuting Pauli strings $\{ \hat{P}_n \}_{n=0}^{N-1}$, it is called a \emph{stabilizer state}. These $N$ operators generate the \emph{stabilizer group} of $\ket{\Psi}$, denoted $\mathcal{S} = \langle \{ \hat{P}_n \}_{n=0}^{N-1} \rangle$. Given a stabilizer state $\ket{\Psi}$ with stabilizer group $\mathcal{S}$, a \emph{stabilizer Hamiltonian} is any linear combination of stabilizer elements $\hat{H}_{\rm stab} = \sum_{ \hat{P} \in \mathcal{S}} c_{\hat{P}} \hat{P}$ where $c_{\hat{P}} \in \mathbb{R}$. The stabilizer state is an energy eigenstate $\hat{H}_{\rm stab} \ket{\Psi} = E \ket{\Psi}$, with the energy eigenvalue $E = \sum_{\hat{P} \in \mathcal{S}} c_{\hat{P}}$, and is a ground state if every coupling constant is chosen so that $c_{\hat{P}} \leq 0$. Well known examples of stabilizer Hamiltonians are the cluster state Hamiltonian \cite{Rau-03a} and the toric code Hamiltonian \cite{Kit-03b}. If every element $\hat{P} \in \mathcal{S}$ of a stabilizer group is at least $\ell$-local and $b$-body, then the stabilizer Hamiltonian is also at least $\ell$-local and $b$-body. Since 2-local, 2-body interactions are the most common in nature, stabilizer Hamiltonians can be difficult to realise in practice if there are no 2-local, 2-body stabilizer elements (for instance the toric code stabilizers elements are at least $4$-body). 

\emph{Parent Hamiltonian construction.} We now introduce a notion of factorizablility of Pauli strings on a lattice, which allows us to construct local, few-body parent Hamiltonians for stabilizer states, even if the stabilizer elements are highly nonlocal or multi-body.

\begin{definition} A Pauli string $\hat{P}$ is $(k,\ell,b)$-factorizable if it can be factorized as a product of $k$ non-trivial Pauli strings that are $\ell$-local and $b$-body, i.e.: \begin{equation} \hat{P} = \hat{P}^{(1)} \hat{P}^{(2)} \hdots \hat{P}^{(k)}, \end{equation} where each $\hat{P}^{(i)}  \neq \hat{\mathbb{I}}^{\otimes N}$ is $\ell$-local and $b$-body. \end{definition}
In general, a Pauli string can be factorized in many different ways. For example, on a $D=1$ dimensional chain the Pauli string $\hat{P} = \hat{\sigma}_n^Z \hat{\sigma}_{n+1}^Z$ is $(2,1,1)$-factorizable, since it can be factorized as $\hat{P} = \hat{P}^{(1)} \hat{P}^{(2)}$ where $\hat{P}^{(1)} = \hat{\sigma}_n^Z$ and $\hat{P}^{(2)} = \hat{\sigma}_{n+1}^Z$ are both 1-local and 1-body. However, it is also $(2,2,2)$-factorizable as $\hat{P} = \hat{P}^{(1)} \hat{P}^{(2)}$ where $\hat{P}^{(1)} = -\hat{\sigma}_n^X \hat{\sigma}_{n+1}^X$ and $\hat{P}^{(2)} = \hat{\sigma}_n^Y \hat{\sigma}_{n+1}^Y$ are both 2-local and 2-body. Based on this factorizability property, we present the following parent Hamiltonian construction:

\begin{result} \label{result:main}
Given a stabilizer state $\ket{\Psi}$ with the stabilizer group $\mathcal{S}$, let $\{ \hat{P}_{\alpha} \}_\alpha \subset \mathcal{S}$ be the subset of $(2,\ell,b)$-factorizable stabilizer group elements $\hat{P}_{\alpha} = \hat{P}_{\alpha}^{(1)} \hat{P}_{\alpha}^{(2)}$, which we enumerate with the index $\alpha$. Then $\ket{\Psi}$ is a zero-energy eigenstate of the $\ell$-local, $b$-body parent Hamiltonian: \begin{equation} \hat{H} = \sum_\alpha J_{\alpha} (\hat{P}^{(1)}_{\alpha} - \hat{P}^{(2)}_{\alpha} ) , \label{eq:H} \end{equation} where $J_{\alpha} \in \mathbb{R}$ are arbitrary coupling constants. 
\end{result}
The proof is straightforward: Since $\hat{P}_\alpha = \hat{P}_\alpha^{(1)} \hat{P}_\alpha^{(2)}$ is a $(2,\ell,b)$-factorizable stabilizer element and $\ket{\Psi}$ is a stabilizer state, we have $\hat{P}_\alpha^{(1)} \hat{P}_\alpha^{(2)} \ket{\Psi} = \ket{\Psi}$, where $\hat{P}_\alpha^{(1)}$ and $\hat{P}_\alpha^{(2)}$ are both $\ell$-local and $b$-body. Multiplying both sides by $\hat{P}_\alpha^{(1)}$, and using $\hat{P}_\alpha^{(1)} \hat{P}_\alpha^{(1)} = \hat{\mathbb{I}}^{\otimes N}$ (since any Pauli string squares to the identity operator), gives $\hat{P}_\alpha^{(2)} \ket{\Psi} = \hat{P}_\alpha^{(1)} \ket{\Psi}$, from which it follows that $( \hat{P}_\alpha^{(1)} - \hat{P}_\alpha^{(2)}) \ket{\Psi} = 0$ for all $\alpha$. The state $\ket{\Psi}$ is therefore annihilated by any linear combination $\hat{H} = \sum_\alpha J_\alpha (\hat{P}^{(1)}_\alpha - \hat{P}^{(2)}_\alpha )$ of such terms. \hfill $\square \quad$

Since the parent Hamiltonian Eq. \ref{eq:H} is a linear combination of non-trivial Pauli strings, it is traceless, and generally has both positive and negative eigenvalues. The zero-energy stabilizer eigenstate is typically near the middle of the spectrum, i.e., a QMBS. We emphasise that the Hamiltonian is $\ell$-local and $b$-body, even if the stabilizer elements are more nonlocal or multi-body. 

A large class of parent Hamiltonians for simple stabilizer states, such as product states, or products of Bell pairs, can be easily constructed using Result \ref{result:main}, as shown in the Supplemental Material (SM) \cite{SupMat}). Here, we provide some examples for more complex stabilizer states.

\begin{figure*}
\includegraphics[width=\textwidth]{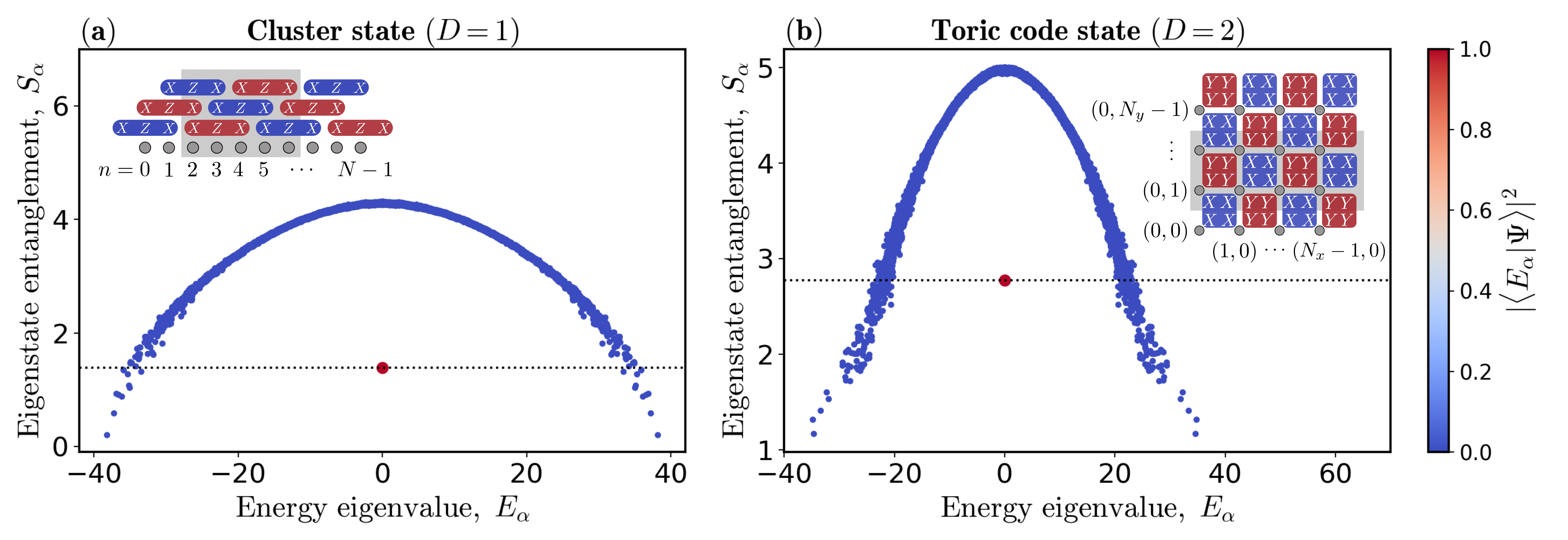}%
\caption{(a) Inset: the generators of the cluster state on the $D=1$ chain. We compute the half-chain entanglement entropy (i.e., the von Neumann entropy for the subsystem highlighted in gray) of each eigenstate of the parent Hamiltonian in Eq. \ref{eq:H_cluster}. The cluster state (red marker) is a zero-energy QMBS (the color of each marker denotes the overlap of the corresponding eigenstate with the cluster state). The dashed black line shows the entanglement entropy $S^{\rm cluster} = 2 \ln 2$ of the cluster state. [Parameters: $N=14$, $\vec{\theta} = (-1,-1,\hdots,-1)$ and $\omega_n$, $\omega'_n$, $J_n$, $J'_n$ are, for each $n$, generated uniformly at random from the range $[0.7,1.3]$.] (b) Inset: the vertex (blue) and plaquette (red) operators on the square lattice. We compute the half-system entanglement entropy (i.e., the von Neumann entropy of the subsystem highlighted in gray) of each eigenstate of the parent Hamiltonian in Eq. \ref{eq:H_toric_general}. The toric code state (red marker) is a zero-energy QMBS. Its entanglement entropy $S^{\rm toric} = N_x \ln 2$ scales with the size of the boundary of the subsystem, rather than its volume. [Parameters: $N_x = N_y = 4$, $\vec{\theta} = (-1,-1,\hdots,-1)$. The Hamiltonian parameters are given in the caption to Fig. \ref{fig:example_toric_level_spacing} in the End Matter.]}  
\label{fig:examples_cluster_toric} 
\end{figure*}  

\emph{Example 1: cluster state ($D=1$).} Consider the cluster state $\ket{\Psi_{\vec{\theta}}^{\rm cluster}}$ on a $D=1$ dimensional chain of $N$ qubits, defined by the stabilizer group $\mathcal{S}_{\vec{\theta}}^{\rm cluster} = \langle \{ \theta_n \hat{\sigma}_{n-1}^X \hat{\sigma}_n^Z \hat{\sigma}_{n+1}^X \}_{n=0}^{N-1} \rangle$. Here $\theta_n \in \{ \pm 1 \}$ and we assume periodic boundary conditions $n + N \equiv n$. Each of the  $N$ independent stabilizer generators  $\theta_n \hat{\sigma}_{n-1}^X \hat{\sigma}_n^Z \hat{\sigma}_{n+1}^X$ [illustrated in the inset to Fig. \ref{fig:examples_cluster_toric}(a)] is 3-local and 3-body, but is $(2,2,2)$-factorizable in two ways: \begin{equation} \theta_n \hat{\sigma}_{n-1}^X \hat{\sigma}_n^Z \hat{\sigma}_{n+1}^X = \left\{ \begin{array}{l} \theta_n (\hat{\sigma}_{n-1}^X) (\hat{\sigma}_n^Z \hat{\sigma}_{n+1}^X) \\  \theta_n (\hat{\sigma}_{n-1}^X \hat{\sigma}_n^Z) (\hat{\sigma}_{n+1}^X) \end{array} \right. , \label{eq:cluster_factors_1} \end{equation} 
Also, the product of nearest neighbour generators is $(2,2,2)$-factorizable: \begin{eqnarray}  ( \theta_n \hat{\sigma}_{n-1}^X && \hat{\sigma}_n^Z \hat{\sigma}_{n+1}^X)( \theta_{n+1} \hat{\sigma}_{n}^X \hat{\sigma}_{n+1}^Z \hat{\sigma}_{n+2}^X) = \nonumber \\ && = \theta_n \theta_{n+1} (\hat{\sigma}_{n-1}^X \hat{\sigma}_n^Y ) ( \hat{\sigma}_{n+1}^Y \hat{\sigma}_{n+2}^X ) , \label{eq:cluster_factors_3} \end{eqnarray} as is the product of next-nearest neighbour generators: \begin{eqnarray} ( \theta_{n-1} \hat{\sigma}_{n-2}^X && \hat{\sigma}_{n-1}^Z \hat{\sigma}_{n}^X)( \theta_{n+1} \hat{\sigma}_{n}^X \hat{\sigma}_{n+1}^Z \hat{\sigma}_{n+2}^X) \nonumber \\ && = \theta_{n-1}  \theta_{n+1} (\hat{\sigma}_{n-2}^X \hat{\sigma}_{n-1}^Z) (\hat{\sigma}_{n+1}^Z \hat{\sigma}_{n+2}^X) . \label{eq:cluster_factors_4} \end{eqnarray}
By Result \ref{result:main}, the $(2,2,2)$-factorizations in Eqs. \ref{eq:cluster_factors_1}-\ref{eq:cluster_factors_4} lead to a 2-local and 2-body parent Hamiltonian for the $D=1$ cluster state of the form:
\begin{eqnarray} \hat{H}^{\rm cluster} &=& \sum_{n} \omega_n (\hat{\sigma}_{n-1}^X - \theta_n \hat{\sigma}_n^Z \hat{\sigma}_{n+1}^X ) \nonumber \\ &+& \sum_n \omega'_n (\hat{\sigma}_{n+1}^X - \theta_n \hat{\sigma}_{n-1}^X \hat{\sigma}_{n}^Z )  \nonumber \\ &+& \sum_n J_n (\hat{\sigma}_{n-1}^X \hat{\sigma}_n^Y - \theta_n \theta_{n+1} \hat{\sigma}_{n+1}^Y \hat{\sigma}_{n+2}^X) \nonumber \\ &+& \sum_n J'_n (\hat{\sigma}_{n-2}^X \hat{\sigma}_{n-1}^Z - \theta_{n-1} \theta_{n+1} \hat{\sigma}_{n+1}^Z \hat{\sigma}_{n+2}^X) , \quad \label{eq:H_cluster} \end{eqnarray} where $\omega_n$, $\omega'_n$, $J_n$ and $J'_n$ are arbitrary real coupling constants. A numerical analysis verifies that $\hat{H}_{\vec{\theta}}^{\rm cluster}$ is nonintegrable (see End Matter, Sec. \ref{app:example_cluster}). However, the cluster state is a nonthermal eigenstate (i.e., a QMBS), since it has zero magic and also a nonthermal value $S^{\rm cluster} = 2 \ln 2$ of half-chain entanglement entropy \cite{Fat-04a}, in contrast to its surrounding thermal states which obey a volume-law scaling $S \propto N$ of entanglement entropy [see Fig. \ref{fig:examples_cluster_toric}(a)]. We note that the freedom to choose the coupling constants and the stabilizer state phases $\theta_n$ gives a large degree of flexibility to construct a parent Hamiltonian $\hat{H}_{\vec{\theta}}^{\rm cluster}$ with a desired property, for instance, $r$-site translation invariance, or with a particular subset of interaction types turned on or off. Longer range 2-body interactions could be added to the Hamiltonian by also finding the $(2,\ell,2)$-factorizable stabilizer elements with $\ell > 2$. 

\emph{Example 2: toric code state ($D=2$).} As another example, we consider one of the most well known many-body stabilizers, the toric code on an $N_x \times N_y$ square lattice with periodic boundary conditions (here, $N_x$ and $N_y$ are both assumed to be even numbers). The stabilizer group of the toric code is generated by a set of ``vertex'' and ``plaquette'' operators, with the vertex operators defined as:
\begin{equation} \hat{\mathcal{O}}^X_{x,y} = \theta_{x,y}\hat{\sigma}^X_{x,y}\hat{\sigma}^X_{x+1,y}\hat{\sigma}^X_{x,y+1}\hat{\sigma}^X_{x+1,y+1} , \quad \theta_{x,y} \in \{ \pm 1 \} , \label{eq:vertex_op} \end{equation} for coordinates $(x,y) \in \mathbb{Z}_{N_x} \times \mathbb{Z}_{N_y}$ such that $x+y$ is even, and the plaquette operators defined as:
\begin{equation} \hat{\mathcal{O}}^Y_{x,y} = \theta_{x,y}\hat{\sigma}^Y_{x,y}\hat{\sigma}^Y_{x+1,y}\hat{\sigma}^Y_{x,y+1}\hat{\sigma}^Y_{x+1,y+1} , \quad \theta_{x,y} \in \{ \pm 1 \} , \label{eq:plaquette_op} \end{equation} at coordinates $(x,y)$ such that $x+y$ is odd [see inset to Fig. \ref{fig:examples_cluster_toric}(b)]. Although there are $N = N_x N_y$ commuting vertex and plaquette operators [one for each site of the lattice], only $N-2$ of them are independent. For a complete set of $N$ independent, commuting generators one can also append the Wilson loop operators $\hat{W}_1 = \theta_{W_1} \prod_{n \in C_1} \hat{\sigma}_n^Z$ and $\hat{W}_2 = \theta_{W_2} \prod_{n \in C_2} \hat{\sigma}_n^Z$, where $C_1$ is a non-contractible ``horizontal'' loop around the torus, $C_2$ is a non-contractible ``vertical'' loop, and $\theta_{W_1}, \theta_{W_2} \in \{ \pm 1 \}$.

Although the vertex and plaquette operators are 4-local and 4-body, they are $(2,2,2)$-factorizable in multiple ways. For example, the vertex operator can be factorized as:
\begin{equation} \hat{\mathcal{O}}^X_{x,y} = \theta_{x,y}( \hat{\sigma}_{x,y}^X \hat{\sigma}^X_{x+1,y} ) (\hat{\sigma}^X_{x,y+1}\hat{\sigma}^X_{x+1,y+1}) , \label{eq:horizontal_factorization} \end{equation}
or as: 
\begin{equation} \hat{\mathcal{O}}^X_{x,y} = \theta_{x,y}( \hat{\sigma}^X_{x,y} \hat{\sigma}^X_{x,y+1}) (\hat{\sigma}^X_{x+1,y} \hat{\sigma}^X_{x+1,y+1} ) , \label{eq:vertical_factorization} \end{equation} which, by Result \ref{result:main}, means that the toric code state is annihilated by the $2$-local operators $\hat{\sigma}^X_{x,y}\hat{\sigma}^X_{x+1,y} - \theta_{x,y} \hat{\sigma}^X_{x,y+1}\hat{\sigma}^X_{x+1,y+1}$ and $\hat{\sigma}^X_{x,y} \hat{\sigma}^X_{x,y+1} - \theta_{x,y}\hat{\sigma}^X_{x+1,y} \hat{\sigma}^X_{x+1,y+1}$. Similarly, the plaquette operator $\hat{\mathcal{O}}^Y_{x,y}$ yields 2-local terms $\hat{\sigma}^Y_{x,y}\hat{\sigma}^Y_{x+1,y} - \theta_{x,y} \hat{\sigma}^Y_{x,y+1}\hat{\sigma}^Y_{x+1,y+1}$ and $\hat{\sigma}^Y_{x,y} \hat{\sigma}^Y_{x,y+1} - \theta_{x,y}\hat{\sigma}^Y_{x+1,y} \hat{\sigma}^Y_{x+1,y+1}$ that annihilate the toric code state for $x+y$ odd. Putting all of these annihilating terms together in a linear combination gives a quantum $XY$ Hamiltonian on the square lattice with correlated interaction strengths on neighbouring parallel bonds. The parent Hamiltonian in its most general form is given in End Matter, Sec. \ref{app:example_toric} but if, for simplicity, we choose $\theta_{x,y} = -1$ for all $(x,y)$, and demand translation invariance in both $x$ and $y$ directions, then we have the parent Hamiltonian: 
\begin{eqnarray} \hat{H}^{\rm toric} = \sum_{x,y} [&J^X& \hat{\sigma}^X_{x,y}\hat{\sigma}^X_{x+1,y} + \tilde{J}^X \hat{\sigma}^X_{x,y}\hat{\sigma}^X_{x,y+1} \nonumber \\ + &J^Y& \hat{\sigma}^Y_{x,y}\hat{\sigma}^Y_{x+1,y} + \tilde{J}^Y \hat{\sigma}^Y_{x,y}\hat{\sigma}^Y_{x,y+1} ] , \label{eq:H_toric} \end{eqnarray} where $J^X$, $J^Y$, $\tilde{J}^X$ and $\tilde{J}^Y$ are real coupling parameters.

The quantum $XY$ Hamiltonian is non-integrable (as verified by an analysis of the level spacing statistics in End Matter, Sec. \ref{app:example_toric}), but has the toric code state as a zero-energy QMBS. The entanglement entropy of the toric code state [corresponding to the bipartion highlighted in gray in the inset to Fig. \ref{fig:examples_cluster_toric}(b)] is $S^{\rm toric} = N_x \ln 2$, which scales with the size of the boundary of the subsystem, rather than the volume-law scaling $S \propto N_x N_y$ of the surrounding thermal eigenstates [see Fig. \ref{fig:examples_cluster_toric}(b)]. Since the Wilson loop operators play no role in constructing the parent Hamiltonian there are, in fact, 4 degenerate toric code QMBS at zero energy, corresponding to the four combinations of the Wilson loop operator phases $\theta_{W_1} \in \{ \pm 1 \}$ and $\theta_{W_2} \in \{ \pm 1 \}$. We note that this construction may shed light on recent results in Ref. \cite{Mia-25a-arxiv}, which discovered QMBS in a $D=2$ dimensional quantum $XY$ model with a $U(1)$-symmetry. Our parent Hamiltonian, however, does not require $U(1)$-symmetry and explicitly identifies the QMBS as the toric code states. Our construction may also help to explain the results of Ref. \cite{Har-25a}, which discovered stabilizer QMBS in a particular lattice gauge theory with similarities to the toric code.

\emph{Volume-law entangled stabilizer QMBS.} Given a bipartition of an $N$-qubit system into subsystems $A$ and $B$, and a stabilizer state $\ket{\Psi}$, its stabilizer group can be decomposed as $\mathcal{S} = \mathcal{S}_A \times \mathcal{S}_B \times \mathcal{S}_{AB}$, where $\mathcal{S}_A$ and $\mathcal{S}_B$ are the subgroups made up of stabilizer elements acting non-trivially only on subsystems $A$ and $B$, and the subgroup $\mathcal{S}_{AB}$ contains all elements not in $\mathcal{S}_A \times \mathcal{S}_B$. With this decomposition, the bipartite entanglement entropy of the stabilizer state is $S = \frac{1}{2} \ln |\mathcal{S}_{AB}|$, where $|\mathcal{S}_{AB}|$ is the number of elements in the subgroup $\mathcal{S}_{AB}$ \cite{Fat-04a}. This suggests an approach to constructing $\ell$-local and $b$-body parent Hamiltonians for volume-law entangled stabilizer QMBS: one should try to find stabilizers with $|\mathcal{S}_{AB}|$ growing exponentially with the volume of subsystem $A$ (or $B$), and containing an extensive number of $(2,\ell,b)$-factorizable elements in the subgroup $\mathcal{S}_{AB}$. This was, in fact, already done for several recent examples of volume-law entangled QMBS, such as the ``rainbow'' Bell pair state \cite{Lan-22a} and the ``antipodal'' Bell pair state \cite{Chi-24a}, although without employing the stabilizer formalism, and without noting that the corresponding QMBS are stabilizer states. In the SM \cite{SupMat} we discuss these recent results in the context of our framework. However, to show how our method can be used to construct parent Hamiltonians for volume-law entangled stabilizer QMBS, we present an example, which we dub the antipodal toric code (ATC) state.

\begin{figure}
\includegraphics[width=\columnwidth]{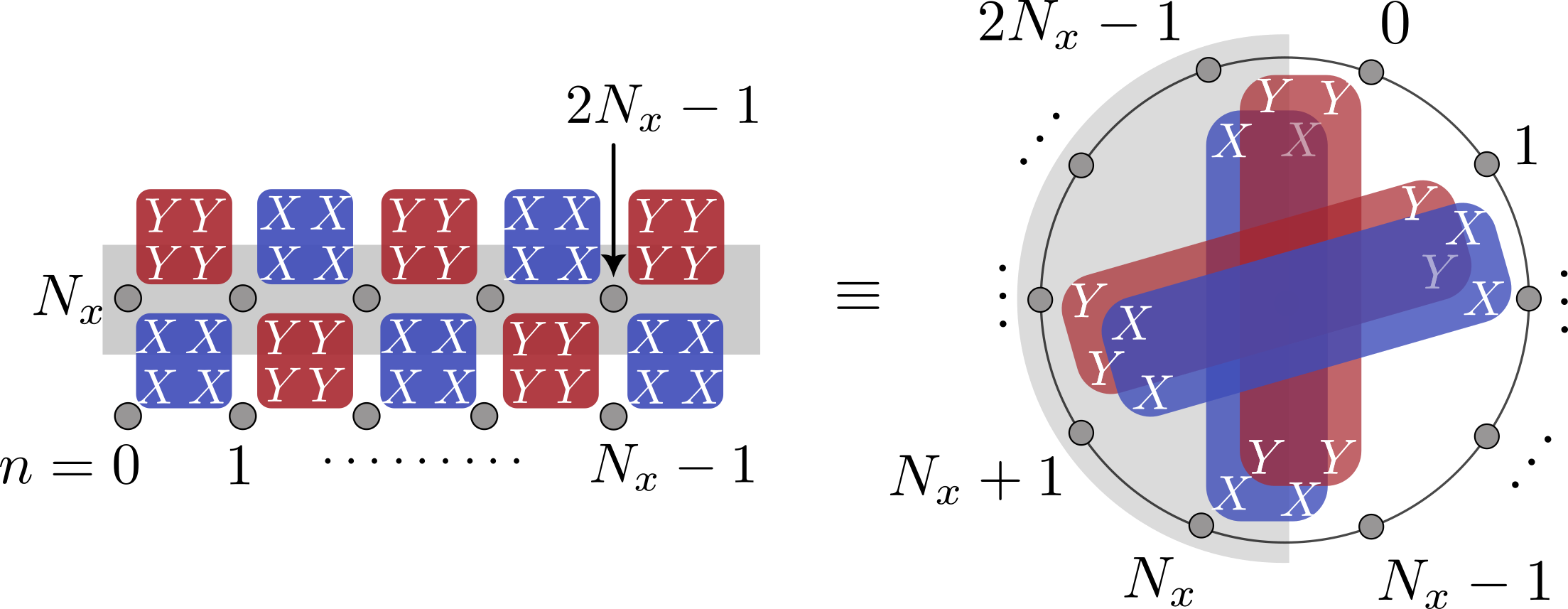}%
\caption{Left: the vertex and plaquette operators for the toric code on a $N_x \times 2$ square lattice, with its coordinates $(x,y)$ mapped onto a linear chain with coordinates $n = 0, 1, \hdots, 2N_x - 1$. Right: an equivalent representation of the operators the circle, showing that they are distributed nonlocally (antipodally). For clarity, only a few vertex/plaquette operators are shown. These vertex and plaquette operators are generators for the stabilizer group of the antipodal toric code (ATC) state.}  
\label{fig:ATC}  
\end{figure}

\emph{Example 3: ATC state.} Consider the previous toric code example, but on a thin torus with $N_y = 2$, and $N_x$ an odd number. To obtain a $D=1$ dimensional chain, we can map the planar coordinates $(x,y)$ to a linear coordinate $n$, via the transformation $n = x + y N_x$, as shown on the left side of Fig. \ref{fig:ATC}. When the $D=1$ dimensional chain is wrapped into a circle, one can see that the vertex and plaquette operators are distributed antipodally, i.e., they are highly nonlocal Pauli strings [see the right side of Fig. \ref{fig:ATC}]. For the bipartition highlighted in gray in Fig. \ref{fig:ATC}, an inspection of the stabilizer group shows that $|\mathcal{S}_{AB}| = 2^{N-2}$, so that the ATC state has entanglement entropy $S = \frac{1}{2}(N - 2) \ln 2$, which is a volume-law scaling (see End Matter, Sec. \ref{app:example_antipodal_toric} for details). Let us now construct a parent Hamiltonian for the ATC state.

For simplicity, we choose the phases $\theta_{n} = +1$ for each vertex (Eq. \ref{eq:vertex_op}) and plaquette (Eq. \ref{eq:plaquette_op}) stabilizer element. With this choice, the product of $\ell$ nearest neighbour vertex operators is a stabilizer element of the form $\hat{\sigma}_n^X \hat{\sigma}_{n+\ell}^X \hat{\sigma}_{n+N_\ell}^X \hat{\sigma}_{n+N_x+\ell}^X$ ($n \in \{0,1,\hdots, N_x -1 \}$), while the product of $\ell$ nearest neighbour plaquette operators is a stabilizer element of the form $\hat{\sigma}_n^Y \hat{\sigma}_{n+\ell}^Y \hat{\sigma}_{n+N_x}^Y \hat{\sigma}_{n+N_x+\ell}^Y$. Their product gives another stabilizer element $\hat{\sigma}_n^Z \hat{\sigma}_{n+\ell}^Z \hat{\sigma}_{n+N_\ell}^Z \hat{\sigma}_{n+N_x+\ell}^Z$. Each of these stabilizer elements is $(2,\ell,2)$-factorizable, so that Result \ref{result:main} can be applied to give the $\ell$-local, $2$-body annihilating terms $\hat{\sigma}_n^X \hat{\sigma}_{n+\ell}^X - \hat{\sigma}_{n+N_x}^X \hat{\sigma}_{n+N_x+\ell}^X$, $\hat{\sigma}_n^Y \hat{\sigma}_{n+\ell}^Y - \hat{\sigma}_{n+N_x}^Y \hat{\sigma}_{n+N_x+\ell}^Y$ and $\hat{\sigma}_n^Z \hat{\sigma}_{n+\ell}^Z - \hat{\sigma}_{n+N_x}^Z \hat{\sigma}_{n+N_x+\ell}^Z$, respectively. If, for simplicity, we insist that the parent Hamiltonian has alternating couplings, then it has the form of an alternating Heisenberg $XYZ$ chain with range $\ell$ interactions: 
\begin{eqnarray} \hat{H}_\ell^{\rm ATC} &=& \sum_{n = 0}^{N-1} (-1)^n (J^X_\ell \hat{\sigma}_n^X \hat{\sigma}_{n+\ell}^X  + J^Y_\ell \hat{\sigma}_n^Y \hat{\sigma}_{n+\ell}^Y + J^Z_\ell \hat{\sigma}_n^Z \hat{\sigma}_{n+\ell}^Z) . \label{eq:H_ATC} \end{eqnarray} 
The ATC state is a zero-energy eigenstate of this Hamiltonian for any interaction range $\ell$, so it is also a zero-energy eigenstate of the sum $ \hat{H}^{\rm ATC} = \sum_{\ell=1}^{N/2} \hat{H}_\ell^{\rm ATC}$. The interaction strengths $J_{\ell}^{\mu}$ can be chosen to decay with the distance $\ell$ with any desired function. For instance, it includes the all-to-all interacting model of Ref. \cite{Muk-25a-arxiv} as a special case. Although the ATC state has volume-law entanglement, it is nonthermal by construction, since it is a stabilizer state. The parent Hamiltonian can be further generalised by noting that, for our thin torus, a Wilson loop operator $\hat{W}_2 = \theta_{W_2} \hat{\sigma}_n^Z \hat{\sigma}_{n+N_x}^Z$ is $(2,1,1)$-factorizable and leads to additional Hamiltonian terms that annihilate the ATC state (see End Matter, Sec. \ref{app:example_antipodal_toric}).  

\emph{Discussion.} Stabilizer states have an efficient classical description in terms of a generating set of Pauli strings, in contrast to thermal states, which are generally believed to lack any efficient classical representation. Nevertheless, stabilizer states can support rich entanglement structures, including multipartite entanglement, long-range entanglement, topological order, and volume-law scaling of entanglement entropy. In this Letter, we introduce a method for constructing physically plausible $\ell$-local, $b$-body Hamiltonians with stabilizer states as QMBS. We have shown that this method enables the construction of many new Hamiltonians with stabilizer states as zero-energy QMBS, and also explains several previously known results in a unified framework. More examples are provided in the SM \cite{SupMat}, including parent Hamiltonians for product states, products of Bell pairs, and volume-law entangled states which we dub the ``rainbow cluster state'' and ``antipodal cluster state''. We also use our framework to show that the PXP model has the stabilizer QMBS recently discovered in Ref. \cite{Iva-25a}.

An advantage of our construction is that the well-developed stabilizer formalism enables efficient computation of various properties of stabilizer states, such as multipartite entanglement \cite{Fat-04a}. Also, stabilizer QMBS can be classified in terms of their complexity, given by the depth of the Clifford circuit that is needed to create them starting from a product stabilizer state (see SM \cite{SupMat}).

Our framework provides wide scope for future work, for example, exploring the quasiparticle excitations on top of stabilizer QMBS, and generalisations to systems of qutrits (or higher dimensional local systems), to parent Hamiltonians with multiple stabilizer QMBS, to stabilizer QMBS at non-zero energies, or to Floquet models with stabilizer QMBS.

\begin{acknowledgments}
I would like to thank L. Coady for invaluable support and encouragement. This publication has emanated from research conducted with the financial support of Taighde \'{E}ireann – Research Ireland under Grant number 22/PATH-S/10812.
\end{acknowledgments} 

\bibliography{/Users/dooleysh/Google_Drive/physics/papers/bibtex_library/refs} 

\clearpage


\vspace{4mm} \begin{center} {\bf \large END MATTER} \end{center}

\section{Further details for Example 1 (the $D=1$ cluster state)} \label{app:example_cluster}

\begin{figure}[b]
\includegraphics[width=\columnwidth]{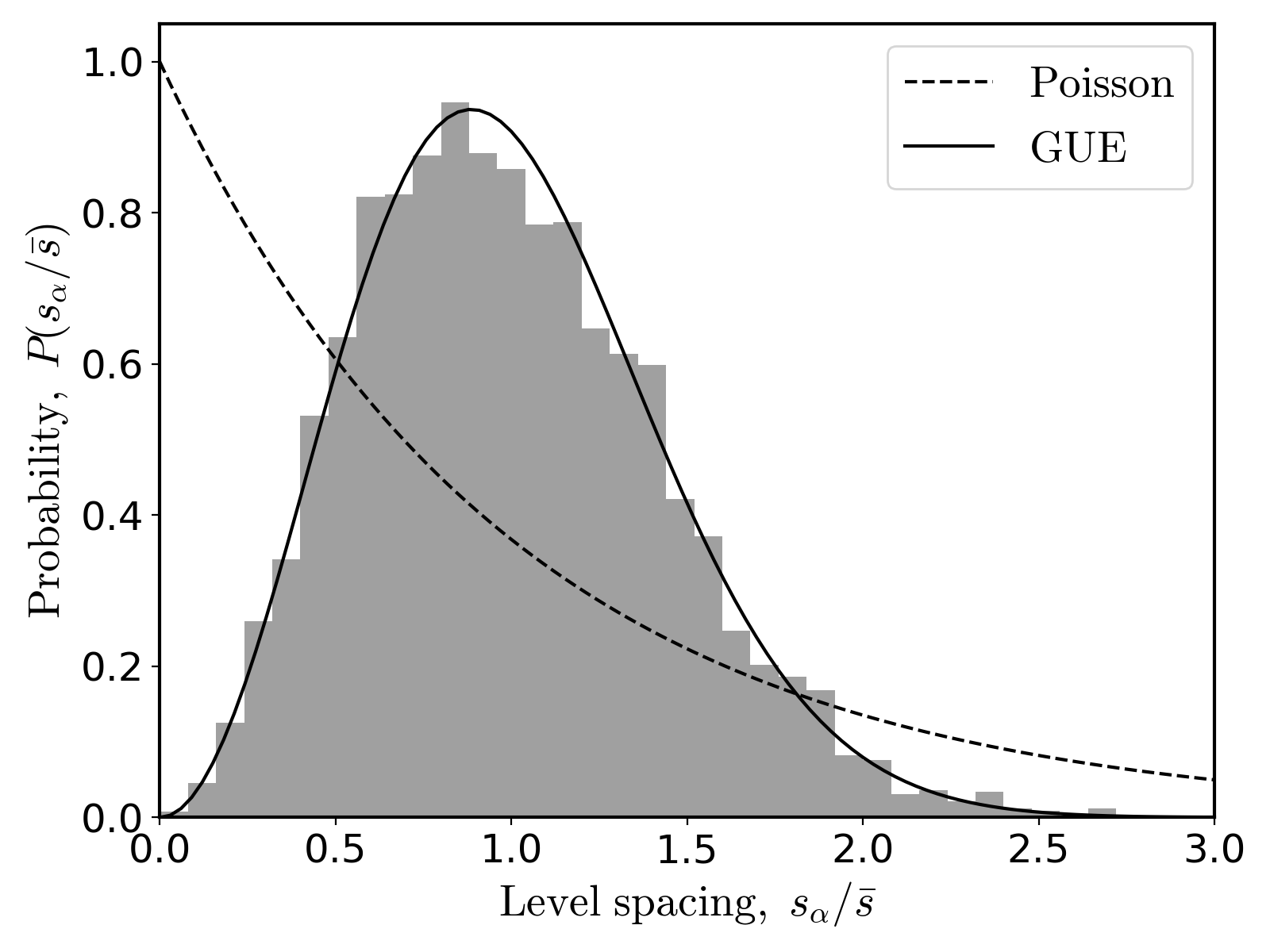}%
\caption{The level spacing statistics for the parent Hamiltonian of the cluster state, given in Eq. \ref{eq:H_cluster}. All parameters are the same as for Fig. \ref{fig:examples_cluster_toric}(a).}  
\label{fig:example_cluster_level_spacing} 
\end{figure}

In Fig. \ref{fig:example_cluster_level_spacing} below we plot the energy level spacing statistics for the parent Hamiltonian of the cluster state, given in Eq. \ref{eq:H_cluster}. Restricting to energy eigenvalues $E_\alpha$ in the middle half of the energy spectrum, we see that level spacings $s_\alpha = E_{\alpha} - E_{\alpha - 1}$ between consecutive energy levels (normalised by the average level spacing $\bar{s}$) follows the same distribution as would be expected for a Hamiltonian drawn from the Gaussian unitary ensemble (GUE). This indicates that the Hamiltonian is nonintegrable and that that Hamiltonian is not time-reversal symmetric \cite{DAl-16}.

The half-chain entanglement entropy of the cluster state is straighforwardly computed by the methods of Ref. \cite{Fat-04a}. Roughly speaking, one can see in the inset to Fig. \ref{fig:examples_cluster_toric}(a) that there are $4$ generators that cross the bipartition (and cannot be constructed as a product of generators acting trivially on each subsystem), so that $|\mathcal{S}_{AB}| = 2^4$. The entanglement entropy of the cluster state is therefore $S = \frac{1}{2} \ln |\mathcal{S}_{AB}| = 2 \ln 2$.

\section{Further details for Example 2 (the toric code state)} \label{app:example_toric}

From the $(2,2,2)$-factorizations of the vertex operator given in Eqs. \ref{eq:horizontal_factorization} and \ref{eq:vertical_factorization}, we know that the toric code state is annihilated by the $2$-local, $2$-body terms $\hat{\sigma}^X_{x,y}\hat{\sigma}^X_{x+1,y} - \theta_{x,y} \hat{\sigma}^X_{x,y+1}\hat{\sigma}^X_{x+1,y+1}$ and $\hat{\sigma}^X_{x,y} \hat{\sigma}^X_{x,y+1} - \theta_{x,y}\hat{\sigma}^X_{x+1,y} \hat{\sigma}^X_{x+1,y+1}$ (for $x+y$ even). Similarly, the plaquette operator $\hat{\mathcal{O}}^Y_{x,y} = \theta_{x,y}\hat{\sigma}^Y_{x,y}\hat{\sigma}^Y_{x+1,y}\hat{\sigma}^Y_{x,y+1}\hat{\sigma}^Y_{x+1,y+1}$ can be $(2,2,2)$-factorized as $\theta_{x,y}(\hat{\sigma}^Y_{x,y}\hat{\sigma}^Y_{x+1,y}) (\hat{\sigma}^Y_{x,y+1}\hat{\sigma}^Y_{x+1,y+1})$ or as $\theta_{x,y} (\hat{\sigma}^Y_{x,y}\hat{\sigma}^Y_{x,y+1})(\hat{\sigma}^Y_{x+1,y}\hat{\sigma}^Y_{x+1,y+1})$, yielding the $2$-local, $2$-body annihilating terms $\hat{\sigma}^Y_{x,y}\hat{\sigma}^Y_{x+1,y} - \theta_{x,y}\hat{\sigma}^Y_{x,y+1}\hat{\sigma}^Y_{x+1,y+1}$ and $\hat{\sigma}^Y_{x,y}\hat{\sigma}^Y_{x,y+1} - \theta_{x,y} \hat{\sigma}^Y_{x+1,y}\hat{\sigma}^Y_{x+1,y+1}$ (for $x+y$ odd). The toric code state is annihilated by any linear combination of these terms:

\begin{eqnarray} \hat{H}^{\rm toric} &=& \sum_{\substack{x,y \\ (x+y)~{\rm even}}} J^X_{x,y} (\hat{\sigma}^X_{x,y}\hat{\sigma}^X_{x+1,y} - \theta_{x,y} \hat{\sigma}^X_{x,y+1}\hat{\sigma}^X_{x+1,y+1} ) \nonumber \\ &+& \sum_{\substack{x,y \\ (x+y)~{\rm even}}} \tilde{J}_{x,y}^X (\hat{\sigma}^X_{x,y} \hat{\sigma}^X_{x,y+1} - \theta_{x,y}\hat{\sigma}^X_{x+1,y} \hat{\sigma}^X_{x+1,y+1}) \} \nonumber \\ &+& \sum_{\substack{x,y \\ (x+y)~{\rm odd}}}  J^Y_{x,y} (\hat{\sigma}^Y_{x,y}\hat{\sigma}^Y_{x+1,y} - \theta_{x,y}\hat{\sigma}^Y_{x,y+1}\hat{\sigma}^Y_{x+1,y+1}) \nonumber \\ &+& \sum_{\substack{x,y \\ (x+y)~{\rm odd}}} \tilde{J}^Y_{x,y} (\hat{\sigma}^Y_{x,y}\hat{\sigma}^Y_{x,y+1} - \theta_{x,y} \hat{\sigma}^Y_{x+1,y}\hat{\sigma}^Y_{x+1,y+1}) . \label{eq:H_toric_general} \end{eqnarray}
This is the quantum $XY$ Hamiltonian on a square lattice, with the restriction that couplings across parallel neighbouring bonds are correlated (if $\theta_{x,y} = -1$) or anticorrelated (if $\theta_{x,y} = 1$). If we assume that $\theta_{x,y} = -1$ for all $x$ and $y$, and that the coupling constants are spatially homogeneous ($J^{X/Y}_{x,y} = J^{X/Y}$ and $\tilde{J}^{X/Y}_{x,y} = \tilde{J}^{X/Y}$) then the Hamiltonian simplifies to Eq. \ref{eq:H_toric}, which is the usual translation invariant quantum $XY$ Hamiltonian on a square lattice.

\begin{figure}[b]
\includegraphics[width=\columnwidth]{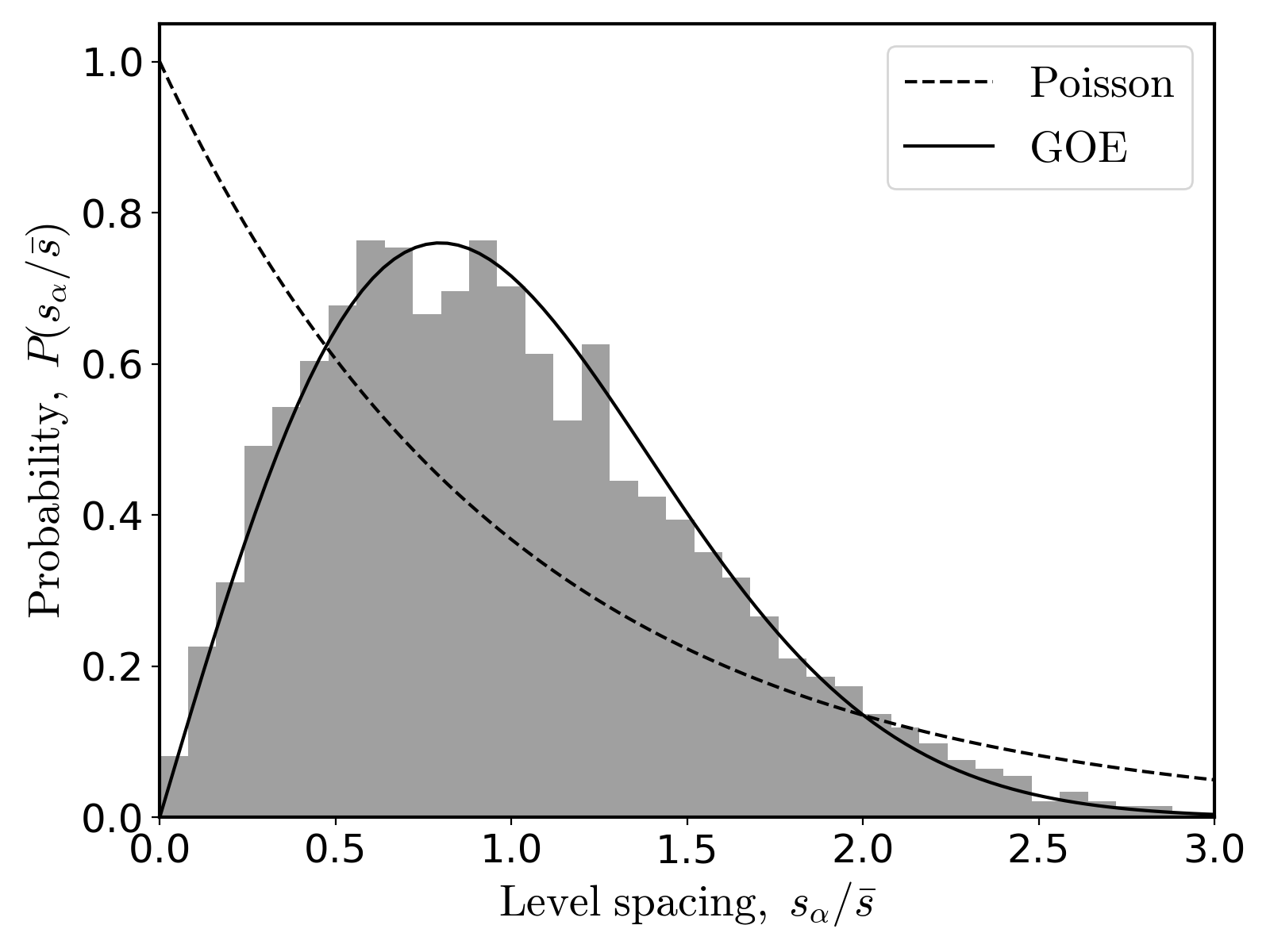}%
\caption{The level spacing statistics for the parent Hamiltonian of the toric code state, given in Eq. \ref{eq:H_toric_general}, in its $\hat{\Pi}^X = \hat{\Pi}^Z = +1$ parity sector. Here, the square lattice has the size $N_x = N_y = 4$, the toric code state is chosen with $\theta_{x,y} = -1$ for all $x$ and $y$, and the coupling constants $J_{x,y}^{X/Y}$ and $\tilde{J}_{x,y}^{X/Y}$ are chosen uniformly at random from the range $[0.7,1.3]$. The same parameters are also used for Fig. \ref{fig:examples_cluster_toric}(b), except that it also shows data in all parity sectors $\hat{\Pi}^X = \pm 1$ and $\hat{\Pi}^Z = \pm 1$.}  
\label{fig:example_toric_level_spacing} 
\end{figure}

It is known that the $D=2$ dimensional quantum $XY$ Hamiltonian is nonintegrable. Here we verify this numerically by computing the level spacing statistics of the Hamiltonian in Eq. \ref{eq:H_toric_general}. Since $\hat{H}^{\rm toric}$ commutes with the $X$-parity operator $\hat{\Pi}^X = \bigotimes_{x,y} \hat{\sigma}_{x,y}^X$ and also the $Z$-parity operator $\hat{\Pi}^Z = \bigotimes_{x,y} \hat{\sigma}_{x,y}^Z$, to obtain sensible level spacing statistics we must restrict to sectors of fixed $X$ and $Z$ parity. In Fig. \ref{fig:example_toric_level_spacing} below we plot the energy level spacing statistics in the $\hat{\Pi}^X = \hat{\Pi}^Z = +1$ parity sector (the sector that includes the toric code states). We see that the level spacing statistics follow the same distribution as would be expected for a Hamiltonian from the Gaussian orthogonal ensemble (GOE). This indicates that the Hamiltonian is nonintegrable and time-reversal symmetric, as expected.

Again, the half-system entanglement entropy of the toric code state can be straighforwardly computed by the methods of Ref. \cite{Fat-04a}. Intuitively, one can see in the inset to Fig. \ref{fig:examples_cluster_toric}(b) that there are $2 N_x$ generators that are cut by the bipartition between the two subsystems (and cannot be constructed as a product of generators acting trivially on each subsystem), so that $|\mathcal{S}_{AB}| = 2^{2N_x}$. The entanglement entropy of the cluster state is therefore $S = \frac{1}{2} \ln |\mathcal{S}_{AB}| = N_x \ln 2$.

\section{Further details for Example 3 (antipodal toric code state)} \label{app:example_antipodal_toric}

In the main text we showed that the ATC state (with stabilizer phases $\theta_n = 1$ for all vertex and plaquette operators) is annihilated by the $\ell$-local, $2$-body terms $\hat{\sigma}_n^X \hat{\sigma}_{n+\ell}^X - \hat{\sigma}_{n+N_x}^X \hat{\sigma}_{n+N_x+\ell}^X$, $\hat{\sigma}_n^Y \hat{\sigma}_{n+\ell}^Y - \hat{\sigma}_{n+N_x}^Y \hat{\sigma}_{n+N_x+\ell}^Y$ and $\hat{\sigma}_n^Z \hat{\sigma}_{n+\ell}^Z - \hat{\sigma}_{n+N_x}^Z \hat{\sigma}_{n+N_x+\ell}^Z$, where  and $n \in \{ 0, 1, \hdots, N_x -1 \}$. The ATC state is therefore annihilated by the parent Hamiltonian obtained by a linear combination of all such terms: \begin{eqnarray} \hat{H}_\ell^{\rm ATC} &=& \sum_{n=0}^{N_x - 1} J_{\ell,n}^X (\hat{\sigma}_n^X \hat{\sigma}_{n+\ell}^X - \hat{\sigma}_{n+N_x}^X \hat{\sigma}_{n+N_x+\ell}^X) \nonumber \\ &+& \sum_{n=0}^{N_x - 1} J_{\ell,n}^Y ( \hat{\sigma}_n^Y \hat{\sigma}_{n+\ell}^Y - \hat{\sigma}_{n+N_x}^Y \hat{\sigma}_{n+N_x+\ell}^Y ) \nonumber \\ &+& \sum_{n=0}^{N_x - 1} J_{\ell,n}^Z ( \hat{\sigma}_n^Z \hat{\sigma}_{n+\ell}^Z - \hat{\sigma}_{n+N_x}^Z \hat{\sigma}_{n+N_x+\ell}^Z ) . \end{eqnarray} Since we have assumed that $N_x$ is an odd number, choosing the coupling constants as $J_{\ell,n}^\mu = (-1)^n J_\ell^\mu$ gives the Hamiltonian for the Heisenber $XYZ$ chain with $\ell$-range alternating couplings given in Eq. \ref{eq:H_ATC}.


There are many more local, few-body terms that annihilate the ATC state, enabling significant generalisation of the parent Hamiltonian. For example, as mentioned in the main text, for our thin torus ($N_y = 2$) the Wilson loop operator $\hat{W}_2 = \theta_{W_2} \hat{\sigma}_0^Z \hat{\sigma}_{N_x}^Z$ is a $(2,1,1)$-factorizable stabilizer element. The product of this operator with the stabilizer element $\hat{\sigma}_0^Z \hat{\sigma}_1^Z \hat{\sigma}_{N_x}^Z \hat{\sigma}_{N_x+1}^Z$ gives another $(2,1,1)$-factorizable stabilizer element $\theta_{W_2} \hat{\sigma}_1^Z \hat{\sigma}_{N_x+1}^Z$. Proceeding iteratively across the chain in this way, we see that $\theta_{W_2} \hat{\sigma}_n^Z \hat{\sigma}_{n+N_x}^Z$ is a $(2,1,1)$-factorizable stabilizer element for the ATC state, for each $n \in \{0,1,\hdots, N_x - 1 \}$. This leads to an additional annihilating parent Hamiltonian: \begin{equation} \hat{H}^Z = \sum_{n=0}^{N_x - 1} h_n^Z ( \hat{\sigma}_n^Z - \theta_{W_2}\hat{\sigma}_{n + N_x}^Z ) , \end{equation} for the ATC state. 

Moreover, the product of any of the stabilizer elements $\theta_{W_2} \hat{\sigma}_n^Z \hat{\sigma}_{n+N_x}^Z$ (obtained above via the Wilson loop operator) with its overlapping plaquette operators $\hat{\sigma}_{n-1}^Y \hat{\sigma}_{n}^Y \hat{\sigma}_{n+N_x-1}^Y \hat{\sigma}_{n+N_x}^Y$ and $\hat{\sigma}_n^Y \hat{\sigma}_{n+1}^Y \hat{\sigma}_{n+N_x}^Y \hat{\sigma}_{n+N_x+1}^Y$ gives the additional $(2,2,2)$-factorizable stabilizer elements $-\theta_{W_2} \hat{\sigma}_{n-1}^Y \hat{\sigma}_{n}^X \hat{\sigma}_{n+N_x-1}^Y \hat{\sigma}_{n+N_x}^X$ and $-\theta_{W_2} \hat{\sigma}_n^X \hat{\sigma}_{n+1}^Y \hat{\sigma}_{n+N_x}^X \hat{\sigma}_{n+N_x+1}^Y$, respectively. Further multiplication by $\ell-1$ neighbouring vertex (or plaquette) operators gives the $(2,\ell,2)$-factorizable stabilizer elements $-\theta_{W_2} \hat{\sigma}_{n}^Y \hat{\sigma}_{n+\ell}^X \hat{\sigma}_{n+N_x}^Y \hat{\sigma}_{n+N_x+\ell}^X$ and $-\theta_{W_2} \hat{\sigma}_n^X \hat{\sigma}_{n+\ell}^Y \hat{\sigma}_{n+N_x}^X \hat{\sigma}_{n+N_x+\ell}^Y$. These lead to the $\ell$-local, $2$-body annihilating terms $\hat{\sigma}_{n}^Y \hat{\sigma}_{n+1}^X + \theta_{W_2} \hat{\sigma}_{n+N_x}^Y \hat{\sigma}_{n+N_x+\ell}^X$ and $\hat{\sigma}_n^X \hat{\sigma}_{n+\ell}^Y + \theta_{W_2} \hat{\sigma}_{n+N_x}^X \hat{\sigma}_{n+N_x+\ell}^Y$ and the annihilating parent Hamiltonians: \begin{eqnarray} \hat{H}^{XY} &=& \sum_{n=0}^{N_x - 1} J_n^{XY} (\hat{\sigma}_n^X \hat{\sigma}_{n+\ell}^Y + \theta_{W_2} \hat{\sigma}_{n+N_x}^X \hat{\sigma}_{n+N_x+\ell}^Y) \nonumber \\ \hat{H}^{YX} &=& \sum_{n=0}^{N_x - 1} J_n^{YX} (\hat{\sigma}_{n}^Y \hat{\sigma}_{n+1}^X + \theta_{W_2} \hat{\sigma}_{n+N_x}^Y \hat{\sigma}_{n+N_x+\ell}^X) . \end{eqnarray} 

The half-system entanglement entropy of the ATC state can be computed by the methods of Ref. \cite{Fat-04a}. One can see in Fig. \ref{fig:ATC} that there are $N = 2 N_x$ vertex and plaquette operators that are cut by the bipartition between the two subsystems. However, as discussed in the main text, for the toric code only $N-2$ of the vertex and plaquette operators are independent. A complete set of stabilizer generators can be obtained by appending the Wilson loop operators $\hat{W}_1 = \theta_{W_1} \bigotimes_{n=0}^{N_x - 1} \hat{\sigma}_n^Z$ (a string of $\hat{\sigma}^Z$ operators in a non-contractable ``horizontal'' loop around the torus) and $\hat{W}_2 = \theta_{W_2} \hat{\sigma}_0^Z \hat{\sigma}_{N_x}^Z$  (a string of $\hat{\sigma}^Z$ operators in a non-contractible ``vertical'' loop around the torus). The operator $\hat{W}_1$ acts nontrivially only on a subsystem $A$, which consists of the qubits labelled $n \in \{0, 1, \hdots, N_x -1\}.$ Since $\hat{\Pi}^Z = \bigotimes_{n=0}^{2N_x-1} \hat{\sigma}_n^Z$ is a stabilizer element, the operator $\hat{\Pi}^Z \hat{W}_1 = \bigotimes_{n=N_x}^{2N_x - 1} \hat{\sigma}_n^Z$ is a also a stabilizer element, which acts nontrivially only on subsystem $B$, consisting of the qubits labelled $n \in \{ N_x, N_x + 1, \hdots, 2N_x -1 \}$. The subgroup $\mathcal{S}_A \times \mathcal{S}_B = \{\hat{\mathbb{I}}^{\otimes N}, \hat{W}_2, \hat{\Pi}_2,  \hat{\Pi}^Z \hat{W}_1 \}$ contains 4 elements, while all other stabilizer elements are in the group $\mathcal{S}_{AB}$, which therefore has the size $|\mathcal{S}_{AB}| = 2^{N-2}$. It follows, by the results of Ref. \cite{Fat-04a}, that the entanglement entropy of the ATC state is $S = \frac{1}{2} \ln |\mathcal{S}_{AB}| = \frac{1}{2} (N - 2) \ln 2$.

\clearpage 

\onecolumngrid
\vspace{4mm} \begin{center} {\bf \large SUPPLEMENTAL MATERIAL} \end{center}

\setcounter{section}{0}

In this Supplemental Material we support the results of the main text by providing some additional technical details and more examples of our parent Hamiltonian construction. In Sec. \ref{app:classifying_stab_states} we describe how stabilizer QMBS can be classified according to the minimum depth of Clifford circuit required to create the stabilizer state. In Sec. \ref{app:product_QMBS} we construct the 2-local, 2-body parent Hamiltonian for the simplest possible stabilizer state: a product state in the computational basis. In Sec. \ref{app:Bell_QMBS} we construct 2-local, 2-body parent Hamiltonians for products of Bell pairs. This includes recently discovered examples of volume-law entangled QMBS, the ``rainbow'' product of Bell pairs (Sec. \ref{app:rainbow_Bell}) \cite{Lan-22a} and the ``antipodal'' product of Bell pairs (Sec. \ref{app:antipodal_Bell}) \cite{Chi-24a}, both of which can be understood within our framework. In Sec. \ref{app:cluster_QMBS} we present two new examples of volume-law entangled QMBS, which we dub the rainbow cluster state and the antipodal cluster state (in analogy with the rainbow product of Bell pairs and the antipodal product of Bell pairs). Finally, in Sec. \ref{app:PXP} we explain that the a recently discovered volume-law entangled eigenstate of the PXP Hamiltonian is a stabilizer state, and can be understood in our framework.

\section{Classifying stabilizer QMBS} \label{app:classifying_stab_states}

Any stabilizer state can be constructed as $\ket{\Psi} = \hat{U} \ket{1}^{\otimes N}$ where $\hat{U}$ is a Clifford circuit and $\hat{\sigma}^Z \ket{\pm 1} = \pm \ket{\pm 1}$ are the qubit computational basis states. One can classify stabilizer states in terms of the minimum depth $d$ of the Clifford circuit $\hat{U}$ required to create the state, where only entangling Clifford gates, such as the controlled-NOT gate contribute to the depth (i.e., single-qubit Clifford gates and non-entangling two-qubit Clifford gates, such as the swap gate, do not contribute to the depth of the circuit).

The simplest stabilizer states are therefore depth $d=0$ product states that are generated by applying only single-qubit Clifford gates to the initial state $\ket{1}^{\otimes N}$. The next simplest stabilizer states are products of Bell pairs, which can be created with depth $d=1$ Clifford circuits by applying a single layer of controlled-NOT gates in parallel (as well as any number of single-qubit Clifford gates and two-qubit swap gates). This set of stabilizer states includes, for example, the ``rainbow'' state of Ref. \cite{Lan-22a}, and the antipodal entangled Bell pair state of Ref. \cite{Chi-24a}. Examples of stabilizer states requiring greater depth are the $D$-dimensional cluster state (minimum depth $d=2D$) and the $D=2$ toric code (minimum depth $d = \mathcal{O}(N)$ \cite{Bra-06b,Shu-25a-arxiv}). This classification provides a convenient way of organising stabilizer QMBS into a hierarchy, depending on their ``complexity''.

\section{Product stabilizer states} \label{app:product_QMBS}

\begin{figure}[b]
\includegraphics[width=\columnwidth]{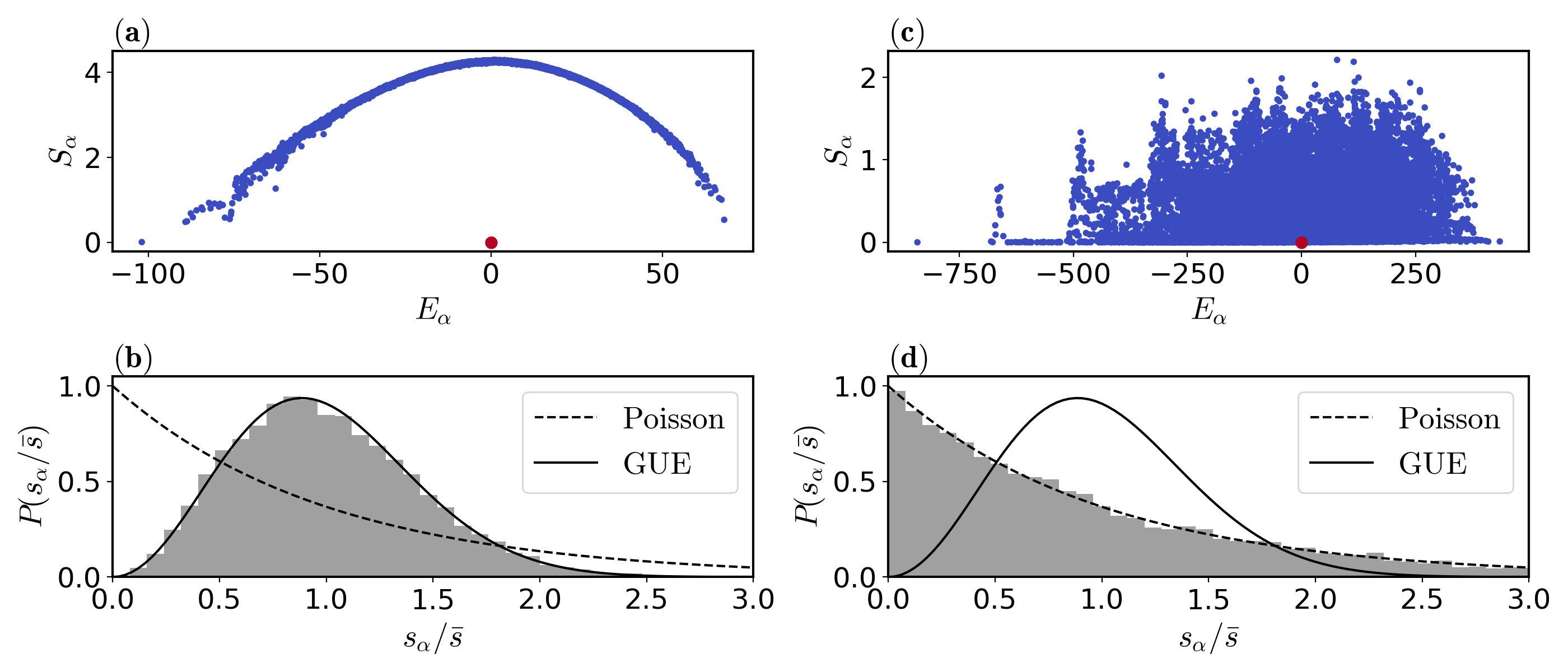}%
\caption{Eigenstate entanglement entropy, and level spacing statistics for two different examples of parent Hamiltonians for the product stabilizer $\ket{1}^{\otimes N}$ ($N=14$). Left column: the Hamiltonian parameters $\omega_n^{\mu,\pm}$, $J_n$, $J'_n$, $\lambda_n^\mu$ are (for each $n$) chosen at random from the interval $[0.7,1.3]$, while $\eta_{n,n'}$, $\gamma_{n,n'}$ and $\xi_{n,n'}$ are chosen at random from the interval $[0.7/N, 1.3/N]$. For this choice of parameters, (a) the mid-spectrum eigenstates follow a volume-law scaling of entanglement entropy, and (b) the Hamiltonian is nonintegrable. However, the product stabilizer QMBS is a zero-entanglement outlier [red marker in (a)]. Right column: the Hamiltonian parameters $\omega_n^{\mu,\pm}$, $J_n$, $J'_n$, $\lambda_n^\mu$ are (for each $n$) chosen at random from the interval $[0.7,1.3]$, $\xi_{n,n'}$ is chosen from the interval $[0.7/N, 1.3/N]$, while each $\eta_{n,n'}$ and $\gamma_{n,n'}$ and are chosen from the interval $[-4, 4]$. This leads to a large amount of disorder in the magnetic field $h_n$. For this choice of parameters, (c) the there are many min-spectrum eigenstates that follow a sub-volume-law scaling of entanglement entropy, and (d) the Hamiltonian is integrable. Again, the product stabilizer is a zero-entanglement eigenstate [red marker in (c)].}  
\label{fig:example_product} 
\end{figure}

Consider the simplest $N$-qubit stabilizer state, which is the (depth $d=0$) product state $\ket{\Psi} = \ket{\theta_0}\ket{\theta_1}\hdots \ket{\theta_{N-1}}$ with the stabilizer group $\mathcal{S}_{\vec{\theta}}^{\rm product} = \langle \{ \theta_n \hat{\sigma}_n^Z \}_{n=0}^{N-1} \rangle$, where $\theta_n \in \{\pm 1\}$ and $\hat{\sigma}^Z \ket{\pm 1} = \pm \ket{\pm 1}$. For simplicity, we focus on the stabilizer state for which $\theta_n = 1$ for all $n$ (i.e., the stabilizer state is $\ket{\Psi} = \ket{1}^{\otimes N}$). In Table \ref{table:product_stabilizer}, we list all of the $(2,2,2)$-factorizable stabilizer elements on a $D=1$ dimensional chain, as well as the corresponding $2$-local, $2$-body Hamiltonian terms that annihilate the product stabilizer state. From Result \ref{result:main}, the corresponding 2-local, 2-body parent hamiltonian is a linear combination of all terms in the right column of Table \ref{table:product_stabilizer}: \begin{eqnarray} \hat{H}^{\rm product} &=& \sum_{n}\sum_{\mu = X,Y,Z}\sum_{\pm} \omega_n^{\mu,\pm} (\hat{\sigma}_n^\mu - \hat{\sigma}_{n \pm 1}^Z \hat{\sigma}_n^\mu) + \sum_n J_n (\hat{\sigma}_n^X \hat{\sigma}_{n+1}^X + \hat{\sigma}_n^Y \hat{\sigma}_{n+1}^Y) + \sum_n J'_n (\hat{\sigma}_n^X \hat{\sigma}_{n+1}^Y - \hat{\sigma}_n^Y \hat{\sigma}_{n+1}^X) \nonumber \\ && + \sum_n \sum_{\mu = X,Y,Z} \lambda_n^\mu (\hat{\sigma}_{n-1}^Z \hat{\sigma}_n^\mu - \hat{\sigma}_n^\mu \hat{\sigma}_{n+1}^Z) + \sum_{n,n'} \eta_{n,n'} (\hat{\sigma}_n^Z - \hat{\sigma}_{n'}^Z)  \nonumber \\ && + \sum_{n,n'} \gamma_{n,n'} (\hat{\sigma}_{n}^Z - \hat{\sigma}_{n'}^Z \hat{\sigma}_{n'+1}^Z) + \sum_{n,n'} \xi_{n,n'} (\hat{\sigma}_n^Z \hat{\sigma}_{n+1}^Z - \hat{\sigma}_{n'}^Z \hat{\sigma}_{n'+1}^Z ) , \label{eq:H_product_messy} \end{eqnarray} where $\omega_n^{\mu,\pm}$, $J_n$, $J'_n$, $\lambda_n^\mu$, $\eta_{n,n'}$, $\gamma_{n,n'}$, $\xi_{n,n'}$ are all arbitrary real numbers. Grouping similar terms together, the Hamiltonian can be somewhat simplified to:
\begin{eqnarray} \hat{H}^{\rm product} &=& \sum_n h_n \hat{\sigma}_n^Z + \sum_n J^Z_n \hat{\sigma}_n^Z \hat{\sigma}_{n+1}^Z + \sum_n J_n (\hat{\sigma}_n^X \hat{\sigma}_{n+1}^X + \hat{\sigma}_n^Y \hat{\sigma}_{n+1}^Y) + \sum_n J'_n (\hat{\sigma}_n^X \hat{\sigma}_{n+1}^Y - \hat{\sigma}_n^Y \hat{\sigma}_{n+1}^X) \nonumber \\ && \qquad + \sum_{n}\sum_{\mu = X,Y,Z}\sum_{\pm} \omega_n^{\mu,\pm} (\hat{\sigma}_n^\mu - \hat{\sigma}_{n \pm 1}^Z \hat{\sigma}_n^\mu) + \sum_n \sum_{\mu = X,Y,Z} \lambda_n^\mu (\hat{\sigma}_{n-1}^Z \hat{\sigma}_n^\mu - \hat{\sigma}_n^\mu \hat{\sigma}_{n+1}^Z)  , \label{eq:H_product_clean} \end{eqnarray} where $h_n \equiv \sum_{n'} (\eta_{n,n'} - \eta_{n',n} + \gamma_{n,n'})$ and $J^Z_n \equiv \sum_{n'} (\xi_{n,n'} - \xi_{n',n} - \gamma_{n',n})$. In Fig. \ref{fig:example_product} we plot the half-chain eigenstate entanglement entropy and level spacing distribution for two different choices of Hamiltonian parameters. In the left column (a,b) the parameters are chosen so that $\hat{H}^{\rm product}$ is nonintegrable, and we clearly see the stabilizer product state as a zero-entropy outlier, i.e., a QMBS. In the right column (c,d), the parameters are chosen to give large disorder in the magnetic field $h_n$, which induces many-body localisation (MBL). In this case, the stabilizer product state is an eigenstate, but is surrounded by many other localised, low-entropy eigenstates.

This construction of a parent Hamiltonian for the product stabilizer state is straighforwardly extended to lattices in higher spatial dimensions $D>1$, and also to longer-range 2-body interactions. We note that a similar parent Hamiltonian for the QMBS $\ket{1}^{\otimes N}$ can also be constructed by the projector embedding method of Shiraishi and Mori \cite{Shi-17a}.

\begin{table}
\begin{center} 
  \begin{tabular}{ c l l }
    \toprule
    \multicolumn{2}{c}{\bf $(2,2,2)$-factorizable stabilizer element} \hspace{5mm} & {\bf $2$-local, $2$-body parent Hamiltonian term}  \\ \midrule 
  $\hat{\sigma}_n^Z$ & $= (\hat{\sigma}_n^Z \hat{\sigma}_{n \pm 1}^\mu)(\hat{\sigma}_{n \pm 1}^\mu)$ & $\hat{\sigma}_{n \pm 1}^\mu - \hat{\sigma}_{n}^Z \hat{\sigma}_{n \pm 1}^\mu$ \\ \midrule
    $\hat{\sigma}_n^Z \hat{\sigma}_{n+1}^Z$ & $= -(\hat{\sigma}_n^X \hat{\sigma}_{n+1}^X) (\hat{\sigma}_n^Y \hat{\sigma}_{n+1}^Y)$ & $\hat{\sigma}_n^X \hat{\sigma}_{n+1}^X + \hat{\sigma}_n^Y \hat{\sigma}_{n+1}^Y$ \\
  & $= (\hat{\sigma}_n^X \hat{\sigma}_{n+1}^Y) (\hat{\sigma}_n^Y \hat{\sigma}_{n+1}^X)$ & $\hat{\sigma}_n^X \hat{\sigma}_{n+1}^Y - \hat{\sigma}_n^Y \hat{\sigma}_{n+1}^X$ \\
    $\hat{\sigma}_{n-1}^Z \hat{\sigma}_{n+1}^Z$ & $= (\hat{\sigma}_{n-1}^Z \hat{\sigma}_{n}^\mu) (\hat{\sigma}_{n}^\mu \hat{\sigma}_{n+1}^Z)$ & $\hat{\sigma}_{n-1}^Z \hat{\sigma}_{n}^\mu - \hat{\sigma}_{n}^\mu \hat{\sigma}_{n+1}^Z $ \\
    $\hat{\sigma}_n^Z \hat{\sigma}_{n'}^Z$ & $ = (\hat{\sigma}_n^Z)(\hat{\sigma}_{n'}^Z)$ & $\hat{\sigma}_n^Z - \hat{\sigma}_{n'}^Z$ \\ \midrule
    $\hat{\sigma}_n^Z \hat{\sigma}_{n'}^Z \hat{\sigma}_{n'+1}^Z$ & $ = (\hat{\sigma}_n^Z) (\hat{\sigma}_{n'}^Z \hat{\sigma}_{n'+1}^Z)$ & $\hat{\sigma}_n^Z - \hat{\sigma}_{n'}^Z \hat{\sigma}_{n'+1}^Z$ \\ \midrule
    $\hat{\sigma}_n^Z \hat{\sigma}_{n+1}^Z \hat{\sigma}_{n'}^Z \hat{\sigma}_{n'+1}^Z$ & $ = (\hat{\sigma}_n^Z \hat{\sigma}_{n+1}^Z ) ( \hat{\sigma}_{n'}^Z \hat{\sigma}_{n'+1}^Z)$ & $\hat{\sigma}_n^Z \hat{\sigma}_{n+1}^Z - \hat{\sigma}_{n'}^Z \hat{\sigma}_{n'+1}^Z$ \\
  \bottomrule
  \end{tabular}
\end{center}
\caption{All $(2,2,2)$-factorizable stabilizer elements of the product stabilizer state $\ket{1}^{\otimes N}$ (with stabilizer group $\mathcal{S}_{(1,1,\hdots,1)}^{\rm product} = \langle \hat{\sigma}_0^{Z}, \hat{\sigma}_1^{Z}, \hdots, \hat{\sigma}_{N-1}^Z \rangle$), and the corresponding 2-local, 2-body operators that annihilate the state $\ket{1}^{\otimes N}$.}
\label{table:product_stabilizer}
\end{table}

\section{Tensor products of Bell pairs} \label{app:Bell_QMBS}

We write the two-qubit Bell states as: \begin{equation} \ket{\psi (\theta, \theta')} = \frac{1}{\sqrt{2}} (\ket{1} \ket{\theta'} + \theta \ket{-1} \ket{-\theta'} ) , \end{equation} where $\theta, \theta' \in \{ \pm 1 \}$ and $\hat{\sigma}^Z \ket{\pm 1} = \pm \ket{\pm 1}$. This is a stabilizer state with the stabilizer group $\mathcal{S} = \langle \{ \theta \hat{\sigma}^X \otimes \hat{\sigma}^X, \theta' \hat{\sigma}^Z \otimes \hat{\sigma}^Z  \} \rangle$. For an $N$ qubit system, with qubits labelled $n \in \{0, 1, \hdots, N-1 \}$, the Bell state of qubits $n$ and $n'$ is denoted $\ket{\psi (\theta_n, \theta_{n'})}_{n,n'}$. A Bell pair can be generated from the product state $\ket{1}\ket{1}$ with one-qubit Clifford unitaries and a single controlled-NOT gate. Therefore, a tensor product of $N/2$ Bell pairs can be generated by applying $N/2$ controlled-NOT gates in parallel, in a depth $d=1$ Clifford circuit.

\begin{figure}
\includegraphics[width=\columnwidth]{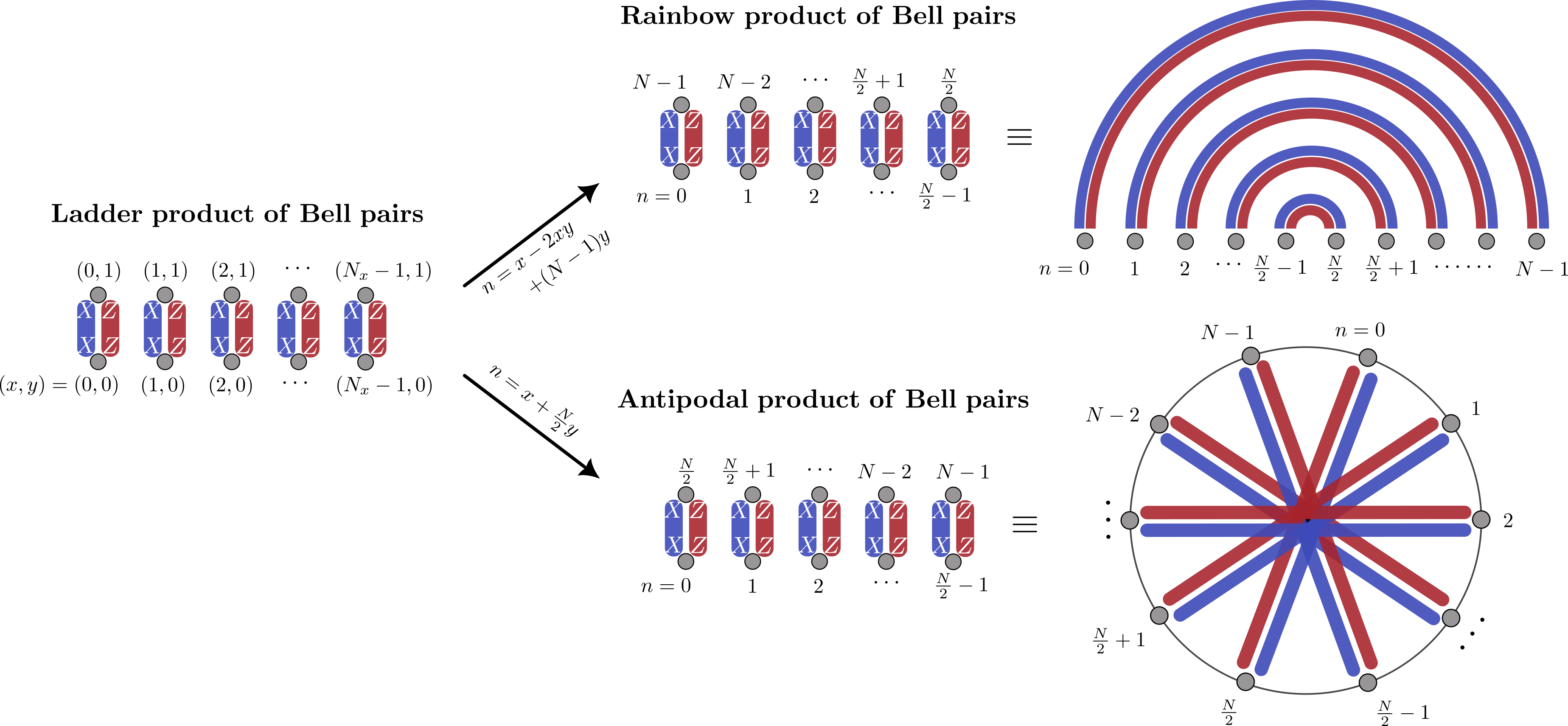}%
\caption{Left: stabilizer generators for the product of Bell pairs on a ladder (i.e., an $N_x \times N_y$ square lattice with $N_y = 2$). Top right: stabilizer generators for the rainbow product of Bell pairs. Bottom right: stabilizer generators for the antipodal product of Bell pairs. The rainbow product is obtained from the ladder product by the mapping $(x,y) \to n = x - 2xy + (N-1)y$ of the ladder to the chain, and the antipodal product is obtained from the ladder product by the mapping $(x,y) \to n = x + \frac{N}{2}y$. Both are nonlocal transformations, in the sense that neighbouring qubits on the ladder are mapped to distant qubits on the chain.}  
\label{fig:products_of_Bell_pairs} 
\end{figure}

Examples of such states have already been considered as QMBS in Ref. \cite{Lan-22a}, called rainbow states, and in Ref. \cite{Chi-24a}, called antipodal Bell pair states, though without employing the stabilizer formalism. Here, we use our framework to construct $2$-local, $2$-body parent Hamiltonians for three different examples of stabilizer states that are tensor products of $N/2$ Bell pairs. The first example we call a ``ladder'' product of Bell pairs (Sec. \ref{app:ladder_Bell}), the second example (Sec. \ref{app:rainbow_Bell}) is the rainbow product of Bell pairs (reproducing some results of Ref. \cite{Lan-22a}) and the third example (Sec. \ref{app:antipodal_Bell} is the antipodal product of Bell pairs (reproducing some results of Ref. \cite{Chi-24a}). Fig. \ref{fig:products_of_Bell_pairs} shows that the stabilizers for the three examples are related by different mappings of the qubits on the ladder to the qubits on the chain.

\subsection{Ladder product of Bell pairs} \label{app:ladder_Bell}

Consider a square lattice in $D=2$ dimensions, with $N_x$ qubits along the $x$-direction and $N_y = 2$ qubits along the $y$-direction [i.e., a ladder; see Fig. \ref{fig:products_of_Bell_pairs} (left)]. We define the ladder product of Bell pairs as the state: \begin{equation} \ket{\Psi^{\rm ladder-BP}} = \bigotimes_{x=0}^{N_x - 1} \ket{\psi (\theta_{x,0}, \theta_{x,1})}_{(x,0),(x,1)} , \label{eq:product_Bell_pairs} \end{equation} where $\ket{\psi (\theta_{x,0}, \theta_{x,1})}_{(x,0),(x,1)}$ is a Bell pair for the two qubits on a rung of the ladder [i.e., at lattice sites $(x,0)$ and $(x,1)$] and $\theta_{x,0} \in \{ \pm 1 \}$, $\theta_{x,1} \in \{ \pm 1 \}$. This is a stabilizer state with the stabilizer group: \begin{equation} \mathcal{S}^{\rm ladder-BP} = \langle \{ \theta_{x,0} \hat{\sigma}_{x,0}^X \hat{\sigma}_{x,1}^X,  \theta_{x,1} \hat{\sigma}_{x,0}^Z \hat{\sigma}_{x,1}^Z \}_{x=0}^{N_x - 1} \rangle . \end{equation} The generators $\theta_{x,0} \hat{\sigma}_{x,0}^X \hat{\sigma}_{x,1}^X$ and $\theta_{x,1} \hat{\sigma}_{x,0}^Z \hat{\sigma}_{x,1}^Z$ are illustrated in Fig. \ref{fig:products_of_Bell_pairs} (left).

In Table \ref{table:product_Bell_pairs} we list all of the $(2,2,2)$-factorizable stabilizer elements of $\mathcal{S}^{\rm ladder-BP}$, along with the corresponding $2$-local, $2$-body Hermitian operators that annihilate the stabilizer state. By Result \ref{result:main}, the 2-local, 2-body parent Hamiltonian for the ladder product of Bell pairs is a linear combination of all terms in the right column of Table \ref{table:product_Bell_pairs}, and is given by $\hat{H}^{\rm ladder-BP} = \hat{H}_h + \hat{H}_\updownarrow + \hat{H}_\leftrightarrow$ where: \begin{eqnarray} \hat{H}_h &=& \sum_{x=0}^{N_x - 1} [ h_{x}^X (\hat{\sigma}_{x,0}^X - \theta_{x,0} \hat{\sigma}_{x,1}^X) + h_{x}^Y (\hat{\sigma}_{x,0}^Y + \theta_{x,0} \theta_{x,1} \hat{\sigma}_{x,1}^Y) + h_{x}^Z (\hat{\sigma}_{x,0}^Z - \theta_{x,1} \hat{\sigma}_{x,1}^Z) ] \label{eq:H_h} \\ \hat{H}_\updownarrow &=& \sum_{x=0}^{N_x-1} [ J_{x}^{XX} (\hat{\sigma}_{x,0}^X \hat{\sigma}_{x,1}^X - \theta_{x,0}\theta_{x+1,0} \hat{\sigma}_{x+1,0}^X \hat{\sigma}_{x+1,1}^X) + J_{x}^{YY} (\hat{\sigma}_{x,0}^Y \hat{\sigma}_{x,1}^Y - \theta_{x,0}\theta_{x,1}\theta_{x+1,0}\theta_{x+1,1} \hat{\sigma}_{x+1,0}^Y \hat{\sigma}_{x+1,1}^Y) \nonumber \\ && \qquad + J_{x}^{ZZ} (\hat{\sigma}_{x,0}^Z \hat{\sigma}_{x,1}^Z - \theta_{x,1}\theta_{x+1,1} \hat{\sigma}_{x+1,0}^Z \hat{\sigma}_{x+1,1}^Z) + J_{x}^{XZ} (\hat{\sigma}_{x,0}^X \hat{\sigma}_{x,1}^X - \theta_{x,0}\theta_{x+1,1} \hat{\sigma}_{x+1,0}^Z \hat{\sigma}_{x+1,1}^Z) \nonumber \\ && \qquad + J_{x}^{ZX} (\hat{\sigma}_{x,0}^Z \hat{\sigma}_{x,1}^Z - \theta_{x,1}\theta_{x+1,0} \hat{\sigma}_{x+1,0}^X \hat{\sigma}_{x+1,1}^X) + J_{x}^{XY} (\hat{\sigma}_{x,0}^X \hat{\sigma}_{x,1}^X + \theta_{x,0}\theta_{x+1,0}\theta_{x+1,1} \hat{\sigma}_{x+1,0}^Y \hat{\sigma}_{x+1,1}^Y) \nonumber \\ && \qquad + J_{x}^{YX} (\hat{\sigma}_{x,0}^Y \hat{\sigma}_{x,1}^Y + \theta_{x,0}\theta_{x,1}\theta_{x+1,0} \hat{\sigma}_{x+1,0}^X \hat{\sigma}_{x+1,1}^X ) + J_{x}^{ZY} ( \hat{\sigma}_{x,0}^Z \hat{\sigma}_{x,1}^Z + \theta_{x,1}\theta_{x+1,0} \theta_{x+1,1} \hat{\sigma}_{x+1,0}^Y \hat{\sigma}_{x+1,1}^Y) \nonumber \\ && \qquad + J_{x}^{YZ} (\hat{\sigma}_{x,0}^Y \hat{\sigma}_{x,1}^Y + \theta_{x,0}\theta_{x,1}\theta_{x+1,1} \hat{\sigma}_{x+1,0}^Z \hat{\sigma}_{x+1,1}^Z) ] \label{eq:H_updown} \\ \hat{H}_{\leftrightarrow} &=& \sum_{x=0}^{N_x-1} [ \tilde{J}_{x}^{XX} (\hat{\sigma}_{x,0}^X \hat{\sigma}_{x+1,0}^X - \theta_{x,0}\theta_{x+1,0} \hat{\sigma}_{x,1}^X \hat{\sigma}_{x+1,1}^X) + \tilde{J}_{x}^{YY} (\hat{\sigma}_{x,0}^Y \hat{\sigma}_{x+1,0}^Y - \theta_{x,0}\theta_{x,1}\theta_{x,1}\theta_{x+1,1} \hat{\sigma}_{x+1,1}^Y \hat{\sigma}_{x+1,1}^Y) \nonumber \\ && \qquad + \tilde{J}_{x}^{ZZ} (\hat{\sigma}_{x,0}^Z \hat{\sigma}_{x+1,0}^Z - \theta_{x,1}\theta_{x,1} \hat{\sigma}_{x+1,1}^Z \hat{\sigma}_{x+1,1}^Z) + \tilde{J}_{x}^{XZ} (\hat{\sigma}_{x,0}^X \hat{\sigma}_{x+1,0}^Z - \theta_{x,0}\theta_{x+1,1} \hat{\sigma}_{x,1}^X \hat{\sigma}_{x+1,1}^Z) \nonumber \\ && \qquad + \tilde{J}_{x}^{ZX} (\hat{\sigma}_{x,0}^Z \hat{\sigma}_{x+1,0}^X - \theta_{x,1}\theta_{x+1,0} \hat{\sigma}_{x,1}^Z \hat{\sigma}_{x+1,1}^X) + \tilde{J}_{x}^{XY} (\hat{\sigma}_{x,0}^X \hat{\sigma}_{x+1,0}^Y + \theta_{x,0}\theta_{x+1,0}\theta_{x+1,1} \hat{\sigma}_{x,1}^X \hat{\sigma}_{x+1,1}^Y) \nonumber \\ && \qquad + \tilde{J}_{x}^{YX} (\hat{\sigma}_{x,0}^Y \hat{\sigma}_{x+1,0}^X + \theta_{x,0}\theta_{x,1}\theta_{x+1,0} \hat{\sigma}_{x,1}^Y \hat{\sigma}_{x+1,1}^X ) + \tilde{J}_{x}^{ZY} ( \hat{\sigma}_{x,0}^Z \hat{\sigma}_{x+1,0}^Y + \theta_{x,1}\theta_{x+1,0} \theta_{x+1,1} \hat{\sigma}_{x,1}^Z \hat{\sigma}_{x+1,1}^Y) \nonumber \\ && \qquad + \tilde{J}_{x}^{YZ} (\hat{\sigma}_{x,0}^Y \hat{\sigma}_{x+1,0}^Z + \theta_{x,0}\theta_{x,1}\theta_{x+1,1} \hat{\sigma}_{x,1}^Y \hat{\sigma}_{x+1,1}^Z) ] . \label{eq:H_leftright} \end{eqnarray} Here, $\hat{H}_h$ describes 1-body magnetic field terms, $\hat{H}_\updownarrow$ describes 2-local, 2-body interactions in the $y$-direction [i.e., between rails of the ladder in Fig. \ref{fig:products_of_Bell_pairs}(left)] and $\hat{H}_\leftrightarrow$ describes 2-local, 2-body interactions in the $x$-direction (i.e., between rungs of the ladder). 

This representation of the parent Hamiltonian, although clearly a linear combination of the annihilating terms in the right column of Table \ref{table:product_Bell_pairs}, is quite complicated. A much simpler representation is possible if we assume that $\theta_{x,y} = 1$ for all $(x,y)$. Then we can write the parent Hamiltonian as $\hat{H}^{\rm ladder-BP} = \hat{H}_0 + \hat{H}_1 + \hat{H}_\updownarrow$, where: \begin{eqnarray} \hat{H}_0 &=& \sum_{x=0}^{N_x - 1} \sum_{\mu \in \{ X,Y,Z \}} h_{x}^\mu \hat{\sigma}_{x,0}^\mu + \sum_{x=0}^{N_x - 1} \sum_{\mu,\nu \in \{ X,Y,Z \}} \tilde{J}_{x}^{\mu\nu} \hat{\sigma}_{x,0}^\mu \hat{\sigma}_{x+1,0}^\nu  , \label{eq:H_0} \end{eqnarray} is a 2-local, 2-body Hamiltonian along the bottom rail of the ladder ($y=0$), and \begin{eqnarray} \hat{H}_1 &=& - \mathcal{T}_y \hat{H}_0^* \mathcal{T}_y ,  \end{eqnarray} is the Hamiltonian along the top rail of the ladder ($y=1$). Here, $\mathcal{T}_y$ is the one-site translation operator in the $y$-direction (we assume periodic boundary conditions) and the complex conjugation is taken with respect to the computational basis. We note that \emph{any} traceless 2-local, 2-body Hamiltonian on an $N$ qubit chain can be written in the form of $\hat{H}_0$ in Eq. \ref{eq:H_0}. The Hamiltonian $\hat{H}_0$ on the bottom rail, and the Hamiltonian $\hat{H}_1 = - \mathcal{T}_y \hat{H}_0^* \mathcal{T}_y$ on the top rail are coupled by the interaction $\hat{H}_{\updownarrow}$ given in Eq. \ref{eq:H_updown}.

\begin{table}
\begin{center} 
  \begin{tabular}{ l l l }
    \toprule
    \multicolumn{2}{c}{\bf $(2,2,2)$-factorizable stabilizer element} \hspace{5mm} & {\bf $2$-local, $2$-body parent}  \\ \multicolumn{2}{c}{}  & {\bf Hamiltonian term}  \\ \midrule 
  $\theta_{x,0} \hat{\sigma}_{x,0}^X \hat{\sigma}_{x,1}^X$ & $= \theta_{x,0} (\hat{\sigma}_{x,0}^X) (\hat{\sigma}_{x,1}^X)$ & $\hat{\sigma}_{x,0}^X - \theta_{x,0} \hat{\sigma}_{x,1}^X $ \\ $\theta_{x,1} \hat{\sigma}_{x,0}^Z \hat{\sigma}_{x,1}^Z$ & $= \theta_{x,1} (\hat{\sigma}_{x,0}^Z) (\hat{\sigma}_{x,1}^Z)$ & $\hat{\sigma}_{x,0}^Z - \theta_{x,1} \hat{\sigma}_{x,1}^Z $ \\ $-\theta_{x,0}\theta_{x,1} \hat{\sigma}_{x,0}^Y \hat{\sigma}_{x,1}^Y$ & $= -\theta_{x,0}\theta_{x,0} (\hat{\sigma}_{x,0}^Y) (\hat{\sigma}_{x,1}^Y)$ & $\hat{\sigma}_{x,0}^Y + \theta_{x,0}\theta_{x,1} \hat{\sigma}_{x,1}^Y $ \\ \midrule
    $\theta_{x,0}\theta_{x+1,0} \hat{\sigma}_{x,0}^X \hat{\sigma}_{x,1}^X \hat{\sigma}_{x+1,0}^X \hat{\sigma}_{x+1,1}^X$ & $= \theta_{x,0}\theta_{x+1,0} (\hat{\sigma}_{x,0}^X \hat{\sigma}_{x,1}^X )( \hat{\sigma}_{x+1,0}^X \hat{\sigma}_{x+1,1}^X)$ & $\hat{\sigma}_{x,0}^X \hat{\sigma}_{x,1}^X - \theta_{x,0}\theta_{x+1,0} \hat{\sigma}_{x+1,0}^X \hat{\sigma}_{x+1,1}^X $ \\
  & $= \theta_{x,0}\theta_{x+1,0} (\hat{\sigma}_{x,0}^X  \hat{\sigma}_{x+1,0}^X) (\hat{\sigma}_{x,1}^X \hat{\sigma}_{x+1,1}^X)$ & $\hat{\sigma}_{x,0}^X  \hat{\sigma}_{x+1,0}^X - \theta_{x,0}\theta_{x+1,0} \hat{\sigma}_{x,1}^X \hat{\sigma}_{x+1,1}^X$ \\ $\theta_{x,1}\theta_{x+1,1} \hat{\sigma}_{x,0}^Z \hat{\sigma}_{x,1}^Z \hat{\sigma}_{x+1,0}^Z \hat{\sigma}_{x+1,1}^Z$ & $= \theta_{x,1}\theta_{x+1,1} (\hat{\sigma}_{x,0}^Z \hat{\sigma}_{x,1}^Z )( \hat{\sigma}_{x+1,0}^Z \hat{\sigma}_{x+1,1}^Z)$ & $\hat{\sigma}_{x,0}^Z \hat{\sigma}_{x,1}^Z - \theta_{x,1}\theta_{x+1,1} \hat{\sigma}_{x+1,0}^Z \hat{\sigma}_{x+1,1}^Z $ \\
  & $= \theta_{x,1}\theta_{x+1,1} (\hat{\sigma}_{x,0}^Z  \hat{\sigma}_{x+1,0}^Z) (\hat{\sigma}_{x,1}^Z \hat{\sigma}_{x+1,1}^Z)$ & $\hat{\sigma}_{x,0}^Z  \hat{\sigma}_{x+1,0}^Z - \theta_{x,1}\theta_{x+1,1} \hat{\sigma}_{x,1}^Z \hat{\sigma}_{x+1,1}^Z$ \\ $\theta_{x,0}\theta_{x+1,0} \theta_{x,1}\theta_{x+1,1} \hat{\sigma}_{x,0}^Y \hat{\sigma}_{x,1}^Y \hat{\sigma}_{x+1,0}^Y \hat{\sigma}_{x+1,1}^Y$ & $= \theta_{x,0}\theta_{x+1,0} \theta_{x,1}\theta_{x+1,1} (\hat{\sigma}_{x,0}^Y \hat{\sigma}_{x,1}^Y )( \hat{\sigma}_{x+1,0}^Y \hat{\sigma}_{x+1,1}^Y)$ & $\hat{\sigma}_{x,0}^Y \hat{\sigma}_{x,1}^Y - \theta_{x,0}\theta_{x+1,0} \theta_{x,1}\theta_{x+1,1} \hat{\sigma}_{x+1,0}^Y \hat{\sigma}_{x+1,1}^Y $ \\
  & $= \theta_{x,0}\theta_{x+1,0} \theta_{x,1}\theta_{x+1,1} (\hat{\sigma}_{x,0}^Y  \hat{\sigma}_{x+1,0}^Y) (\hat{\sigma}_{x,1}^Y \hat{\sigma}_{x+1,1}^Y)$ & $\hat{\sigma}_{x,0}^Y  \hat{\sigma}_{x+1,0}^Y - \theta_{x,0}\theta_{x+1,0} \theta_{x,1}\theta_{x+1,1} \hat{\sigma}_{x,1}^Y \hat{\sigma}_{x+1,1}^Y$ \\ \midrule
   $\theta_{x,0}\theta_{x+1,1} \hat{\sigma}_{x,0}^X \hat{\sigma}_{x,1}^X \hat{\sigma}_{x+1,0}^Z \hat{\sigma}_{x+1,1}^Z$ & $= \theta_{x,0}\theta_{x+1,1} (\hat{\sigma}_{x,0}^X \hat{\sigma}_{x,1}^X )( \hat{\sigma}_{x+1,0}^Z \hat{\sigma}_{x+1,1}^Z)$ & $\hat{\sigma}_{x,0}^X \hat{\sigma}_{x,1}^X - \theta_{x,0}\theta_{x+1,1} \hat{\sigma}_{x+1,0}^Z \hat{\sigma}_{x+1,1}^Z $ \\
  & $= \theta_{x,0}\theta_{x+1,1} (\hat{\sigma}_{x,0}^X  \hat{\sigma}_{x+1,0}^Z) (\hat{\sigma}_{x,1}^X \hat{\sigma}_{x+1,1}^Z)$ & $\hat{\sigma}_{x,0}^X  \hat{\sigma}_{x+1,0}^Z - \theta_{x,0}\theta_{x+1,1} \hat{\sigma}_{x,1}^X \hat{\sigma}_{x+1,1}^Z$   \\ $\theta_{x,1}\theta_{x+1,0} \hat{\sigma}_{x,0}^Z \hat{\sigma}_{x,1}^Z \hat{\sigma}_{x+1,0}^X \hat{\sigma}_{x+1,1}^X$ & $= \theta_{x,1}\theta_{x+1,0} (\hat{\sigma}_{x,0}^Z \hat{\sigma}_{x,1}^Z )( \hat{\sigma}_{x+1,0}^X \hat{\sigma}_{x+1,1}^X)$ & $\hat{\sigma}_{x,0}^Z \hat{\sigma}_{x,1}^Z - \theta_{x,1}\theta_{x+1,0} \hat{\sigma}_{x+1,0}^X \hat{\sigma}_{x+1,1}^X $ \\
  & $= \theta_{x,1}\theta_{x+1,0} (\hat{\sigma}_{x,0}^Z  \hat{\sigma}_{x+1,0}^X) (\hat{\sigma}_{x,1}^Z \hat{\sigma}_{x+1,1}^X)$ & $\hat{\sigma}_{x,0}^Z  \hat{\sigma}_{x+1,0}^X - \theta_{x,1}\theta_{x+1,0} \hat{\sigma}_{x,1}^Z \hat{\sigma}_{x+1,1}^X$  \\ $- \theta_{x,0}\theta_{x+1,0}\theta_{x+1,1} \hat{\sigma}_{x,0}^X \hat{\sigma}_{x,1}^X \hat{\sigma}_{x+1,0}^Y \hat{\sigma}_{x+1,1}^Y$ & $= -\theta_{x,0}\theta_{x+1,0}\theta_{x+1,1} (\hat{\sigma}_{x,0}^X \hat{\sigma}_{x,1}^X) (\hat{\sigma}_{x+1,0}^Y \hat{\sigma}_{x+1,1}^Y )$ & $\hat{\sigma}_{x,0}^X \hat{\sigma}_{x,1}^X + \theta_{x,0}\theta_{x+1,0}\theta_{x+1,1} \hat{\sigma}_{x+1,0}^Y \hat{\sigma}_{x+1,1}^Y$ \\
  & $= - \theta_{x,0}\theta_{x+1,0}\theta_{x+1,1} (\hat{\sigma}_{x,0}^X \hat{\sigma}_{x+1,0}^Y ) ( \hat{\sigma}_{x,1}^X \hat{\sigma}_{x+1,1}^Y)$ & $\hat{\sigma}_{x,0}^X \hat{\sigma}_{x+1,0}^Y + \theta_{x,0}\theta_{x+1,0}\theta_{x+1,1} \hat{\sigma}_{x,1}^X \hat{\sigma}_{x+1,1}^Y$  \\ $- \theta_{x,0}\theta_{x,1}\theta_{x+1,0} \hat{\sigma}_{x,0}^Y \hat{\sigma}_{x,1}^Y \hat{\sigma}_{x+1,0}^X \hat{\sigma}_{x+1,1}^X$ & $= -\theta_{x,0}\theta_{x,1}\theta_{x+1,0} (\hat{\sigma}_{x,0}^Y \hat{\sigma}_{x,1}^Y) (\hat{\sigma}_{x+1,0}^X \hat{\sigma}_{x+1,1}^X )$ & $\hat{\sigma}_{x,0}^Y \hat{\sigma}_{x,1}^Y + \theta_{x,0}\theta_{x,1}\theta_{x+1,0} \hat{\sigma}_{x+1,0}^X \hat{\sigma}_{x+1,1}^X$ \\
  & $= - \theta_{x,0}\theta_{x,1}\theta_{x+1,0} (\hat{\sigma}_{x,0}^Y \hat{\sigma}_{x+1,0}^X ) ( \hat{\sigma}_{x,1}^Y \hat{\sigma}_{x+1,1}^X)$ & $\hat{\sigma}_{x,0}^Y \hat{\sigma}_{x+1,0}^X + \theta_{x,0}\theta_{x,1}\theta_{x+1,0} \hat{\sigma}_{x,1}^Y \hat{\sigma}_{x+1,1}^X$  \\  $- \theta_{x,1}\theta_{x+1,0}\theta_{x+1,1} \hat{\sigma}_{x,0}^Z \hat{\sigma}_{x,1}^Z \hat{\sigma}_{x+1,0}^Y \hat{\sigma}_{x+1,1}^Y$ & $= -\theta_{x,1}\theta_{x+1,0}\theta_{x+1,1} (\hat{\sigma}_{x,0}^Z \hat{\sigma}_{x,1}^Z) (\hat{\sigma}_{x+1,0}^Y \hat{\sigma}_{x+1,1}^Y )$ & $\hat{\sigma}_{x,0}^Z \hat{\sigma}_{x,1}^Z + \theta_{x,1}\theta_{x+1,0}\theta_{x+1,1} \hat{\sigma}_{x+1,0}^Y \hat{\sigma}_{x+1,1}^Y$ \\
  & $= - \theta_{x,1}\theta_{x+1,0}\theta_{x+1,1} (\hat{\sigma}_{x,0}^Z \hat{\sigma}_{x+1,0}^Y ) ( \hat{\sigma}_{x,1}^Z \hat{\sigma}_{x+1,1}^Y)$ & $\hat{\sigma}_{x,0}^Z \hat{\sigma}_{x+1,0}^Y + \theta_{x,1}\theta_{x+1,0}\theta_{x+1,1} \hat{\sigma}_{x,1}^Z \hat{\sigma}_{x+1,1}^Y$  \\ $- \theta_{x,0}\theta_{x,1}\theta_{x+1,1} \hat{\sigma}_{x,0}^Y \hat{\sigma}_{x,1}^Y \hat{\sigma}_{x+1,0}^Z \hat{\sigma}_{x+1,1}^Z$ & $= -\theta_{x,0}\theta_{x,1}\theta_{x+1,1} (\hat{\sigma}_{x,0}^Y \hat{\sigma}_{x,1}^Y) (\hat{\sigma}_{x+1,0}^Z \hat{\sigma}_{x+1,1}^Z )$ & $\hat{\sigma}_{x,0}^Y \hat{\sigma}_{x,1}^Y + \theta_{x,0}\theta_{x,1}\theta_{x+1,1} \hat{\sigma}_{x+1,0}^Z \hat{\sigma}_{x+1,1}^Z$ \\
  & $= - \theta_{x,0}\theta_{x,1}\theta_{x+1,1} (\hat{\sigma}_{x,0}^Y \hat{\sigma}_{x+1,0}^Z ) ( \hat{\sigma}_{x,1}^Y \hat{\sigma}_{x+1,1}^Z)$ & $\hat{\sigma}_{x,0}^Y \hat{\sigma}_{x+1,0}^Z + \theta_{x,0}\theta_{x,1}\theta_{x+1,1} \hat{\sigma}_{x,1}^Y \hat{\sigma}_{x+1,1}^Z$  \\
  \bottomrule
  \end{tabular}
\end{center}
\caption{All $(2,2,2)$-factorizable stabilizer elements of the ladder product of Bell pairs in Eq. \ref{eq:product_Bell_pairs} (with the stabilizer group $\mathcal{S}^{\rm ladder-BP} = \langle \{ \theta_{x,0} \hat{\sigma}_{x,0}^{X} \hat{\sigma}_{x,1}^{X}, \theta_{x,1} \hat{\sigma}_{x,0}^{Z} \hat{\sigma}_{x,1}^{Z} \}_{x=0}^{N_x-1} \rangle$), and the corresponding 2-local, 2-body operators that annihilate the stabilizer state.}
\label{table:product_Bell_pairs}
\end{table}

\subsection{Rainbow product of Bell pairs} \label{app:rainbow_Bell}

Consider a $D=1$ dimensional chain of $N$ qubits, labelled $n \in \{0,1,\hdots, N-1 \}$, where $N$ is assumed to be an even number. On this system, the rainbow product of Bell pairs is the state \cite{Lan-22a}: \begin{equation} \ket{\Psi^{\rm rainbow-BP}} = \bigotimes_{n=0}^{N/2 - 1} \ket{\psi (\theta_n, \theta_{N-1-n})}_{n,N-1-n} , \end{equation} which is a stabilizer state with the stabilizer group: \begin{equation} \mathcal{S}^{\rm rainbow-BP} = \langle \{ \theta_{n} \hat{\sigma}_{n}^X \hat{\sigma}_{N-1-n}^X,  \theta_{N-1-n} \hat{\sigma}_{n}^Z \hat{\sigma}_{N-1-n}^Z \}_{n=0}^{N/2 - 1} \rangle . \end{equation} The stabilizer generators $\theta_{n} \hat{\sigma}_{n}^X \hat{\sigma}_{N-1-n}^X$ and $\theta_{N-1-n} \hat{\sigma}_{n}^Z \hat{\sigma}_{N-1-n}^Z$ are illustrated in Fig. \ref{fig:products_of_Bell_pairs} (top right), which explains the ``rainbow'' part of the name.

In Fig. \ref{fig:products_of_Bell_pairs} (top) we also see that rainbow product of Bell pairs is related to the product of Bell pairs on the ladder through a mapping $n = x - 2xy + (N - 1) y$ of the ladder coordinates $(x,y)$ to the chain coordinate $n$ (where $N = 2N_x$). This is a nonlocal mapping, in the sense that neigbouring qubits at $(x,0)$ and $(x,1)$ on the ladder are mapped to distant qubits $n$ and $N-1-n$ on the chain.

One could find a 2-local, 2-body parent Hamiltonain for the rainbow product of Bell pairs by finding the $(2,2,2)$-factorizable stabilizer elements of $\mathcal{S}^{\rm rainbow-BP}$ and applying our Result \ref{result:main}. However, since we already did this for the ladder product of Bell pairs, and we know that it is related to the rainbow state by the mapping $n = x - 2xy + (N - 1) y$, we can simply apply this mapping to the parent Hamiltonian in Eqs. \ref{eq:H_h}-\ref{eq:H_leftright}. Assuming $\theta_n = 1$ for all $n$, we find the parent Hamiltonian for the rainbow product of Bell pairs:

\begin{equation} \hat{H}^{\rm rainbow-BP} = \hat{H}_0  + \hat{H}_1 + \hat{H}_\updownarrow  , \end{equation} where: \begin{equation}  \hat{H}_0 = \sum_{n=0}^{N/2 - 1} \sum_{\mu \in \{ X,Y,Z \}} h_{n}^\mu \hat{\sigma}_{n}^\mu + \sum_{n=0}^{N/2 - 1} \sum_{\mu,\nu \in \{ X,Y,Z \}} \tilde{J}_{n}^{\mu\nu} \hat{\sigma}_{n}^\mu \hat{\sigma}_{n+1}^\nu , \end{equation} is an arbitrary 2-local, 2-body Hamiltonian on the first $N/2$ qubits and $\hat{H}_1 = - \mathcal{M} \hat{H}_0^* \mathcal{M}$ is the Hamiltonian on the last $N/2$ qubits, where $\mathcal{M}$ is the mirror operator, which maps $n \to N-1-n$. The Hamiltonians $\hat{H}_0$ and $\hat{H}_1$ are coupled by $\hat{H}_\updownarrow$ in Eq. \ref{eq:H_updown}, which is 2-local and 2-body on a chain with periodic boundary conditions if $J_{x}^{\mu\nu} = 0$ for all $x \neq N_x - 1$. This reproduces the main result of Ref. \cite{Lan-22a}.

\subsection{Antipodal Bell pair state} \label{app:antipodal_Bell}

Again, focussing on a $D=1$ dimensional chain with an even number of qubits $N$, the antipodal Bell pair state \cite{Chi-24a} is defined as: \begin{equation} \ket{\Psi^{\rm antipodal-BP}} = \bigotimes_{n=0}^{N/2 - 1} \ket{\psi (\theta_n, \theta_{n+N/2})}_{n,n+N/2} , \label{eq:psi_antipodal_Bell} \end{equation} which is a stabilizer state with the stabilizer group: \begin{equation} \mathcal{S}^{\rm antipodal-BP} = \langle \{ \theta_{n} \hat{\sigma}_{n}^X \hat{\sigma}_{n+N/2}^X,  \theta_{n+N/2} \hat{\sigma}_{n}^Z \hat{\sigma}_{n+N/2}^Z \}_{n=0}^{N/2 - 1} \rangle . \label{eq:antipodal_Bell_stab} \end{equation} The generators $\theta_{n} \hat{\sigma}_{n}^X \hat{\sigma}_{n+N/2}^X$ and $\theta_{n+N/2} \hat{\sigma}_{n}^Z \hat{\sigma}_{n+N/2}^Z$ are illustrated in Fig. \ref{fig:products_of_Bell_pairs} (bottom right), which explains the ``antipodal'' part of the name.

In Fig. \ref{fig:products_of_Bell_pairs} (bottom right) we also plot the generators for the antipodal Bell pair state on the ladder, with the chain position index $n$ mapped to the $(x,y)$ ladder coordinates by the transformation $n = x + y N/2$, where $N = 2N_x$. The transformation is nonlocal, in the sense that neigbouring qubits at $(x,0)$ and $(x,1)$ on the ladder are mapped to distant qubits $n$ and $n+N/2$ on the chain. A parent Hamiltonain for the antipodal product of Bell pairs is obtained by applying the transformation $n = x + y N/2$ to the parent Hamiltonian for the ladder product of Bell pairs obtained in Eqs. \ref{eq:H_h}-\ref{eq:H_leftright}. If we choose the stabilizer phases $\theta_n = 1$ for all $n$, the Hamiltonian parameters $h_x^Y = h^Y$, $\tilde{J}_x^{XY} = \tilde{J}^{XY}$, $\tilde{J}_x^{YX} = \tilde{J}^{YX}$, $\tilde{J}_x^{ZY} = \tilde{J}^{ZY}$, $\tilde{J}_x^{YZ} = \tilde{J}^{YZ}$ to be independent of the coordinate $x$, and all other Hamiltonian parameters to be zero, we find the parent Hamiltonian:
\begin{equation} \hat{H}^{\rm antipodal-BP} = \sum_{n=0}^{N-1} (h^Y \hat{\sigma}_n^Y + \tilde{J}^{XY} \hat{\sigma}_n^X \hat{\sigma}_{n+1}^Y + \tilde{J}^{YX} \hat{\sigma}_n^Y \hat{\sigma}_{n+1}^X + + \tilde{J}^{ZY} \hat{\sigma}_n^Z \hat{\sigma}_{n+1}^Y + \tilde{J}^{YZ} \hat{\sigma}_n^Y \hat{\sigma}_{n+1}^Z  ) , \end{equation} which reproduces Eq. 10 of Ref. \cite{Chi-24a}. The other parent Hamiltonians in Ref. \cite{Chi-24a} can also be reproduced by appropriate choices of the stabilizer phases $\theta_n$ and Hamiltonian parameters in Eqs. \ref{eq:H_h}-\ref{eq:H_leftright}.

\section{Cluster states} \label{app:cluster_QMBS}

In this section we present several examples of stabilizer states based on the cluster state, and we construct their parent Hamiltonians.

\begin{figure}
\includegraphics[width=\columnwidth]{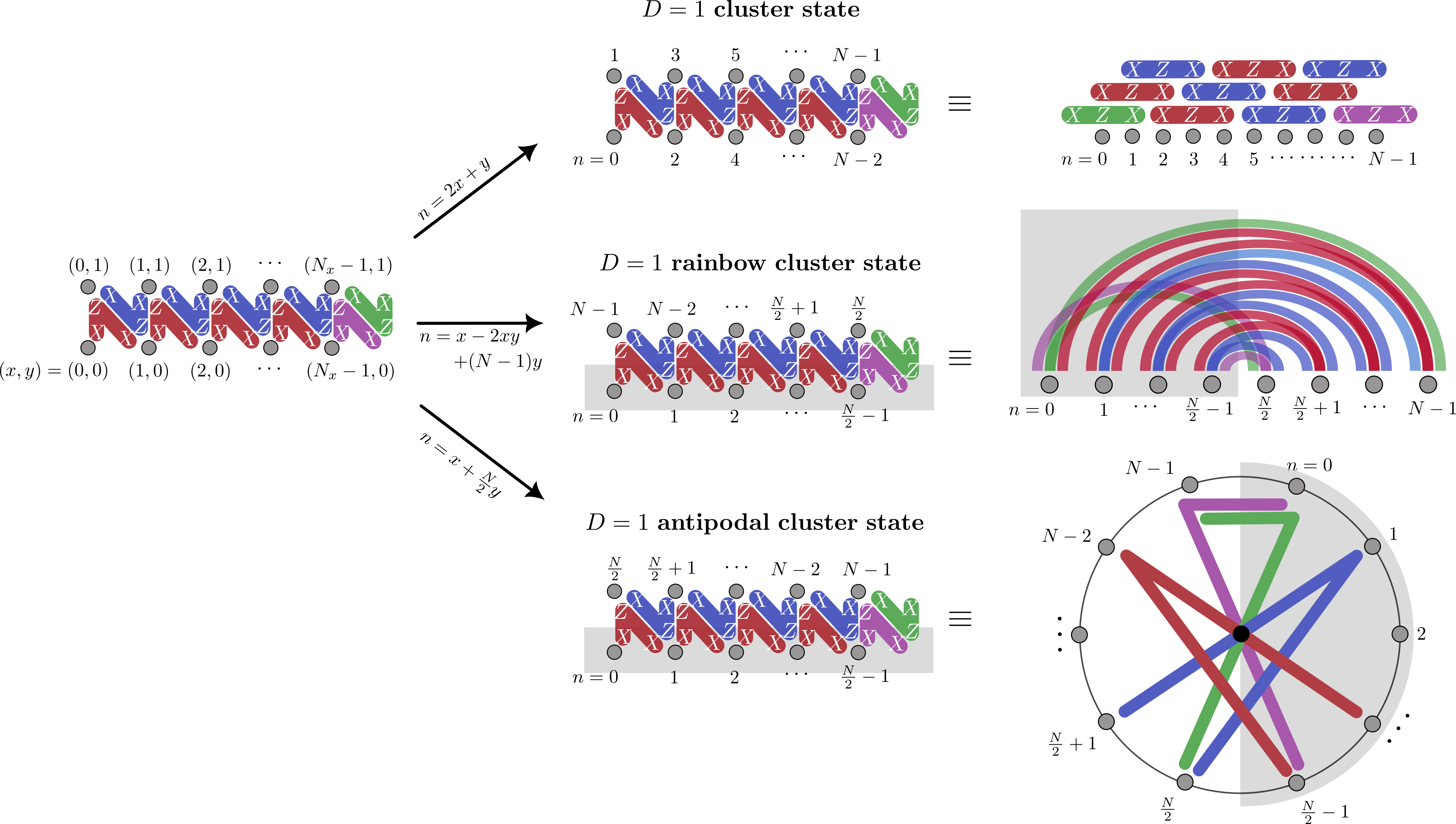}%
\caption{Left: stabilizer generators for the cluster state on a ladder (i.e., an $N_x \times N_y$ square lattice with $N_y = 2$). Top right: stabilizer generators for the standard cluster state on a $D=1$ dimensional chain, obtained by the mapping $(x,y) \to n = 2x + y$ of the ladder coordinates $(x,y)$ to the chain coordinate $n$. Middle right: stabilizer generators for the rainbow cluster state, obtained by the mapping $(x,y) \to n = x - 2xy + (N-1)y$ of the ladder to the chain. Bottom right: stabilizer generators for the antipodal cluster state, obtained by the mapping $(x,y) \to n = x + \frac{N}{2}y$. The transformations to the rainbow and antipodal state are nonlocal, in the sense that neighbouring qubits on the ladder are mapped to distant qubits on the chain, but the mapping to the standard cluster state is local.}  
\label{fig:cluster_schematic} 
\end{figure}

We start by introducing a cluster state on a ladder (i.e., an $N_x \times N_y$ square lattice with $N_y = 2$), defined by the generators [see Fig. \ref{fig:cluster_schematic} (left)]:
\begin{equation} \mathcal{S}^{\rm ladder-cluster} = \langle \{ \theta_{x,0} \hat{\sigma}_{x-1,1}^X \hat{\sigma}_{x,0}^Z \hat{\sigma}_{x,1}^X, \theta_{x,1} \hat{\sigma}_{x,0}^X \hat{\sigma}_{x,1}^Z \hat{\sigma}_{x+1,0}^X  \}_{x=0}^{N_x - 1} \rangle , \end{equation}
where $\theta_{x,y} \in \{ \pm 1 \}$ and $\hat{\sigma}_{x,y}^\mu$ is a Pauli operator at the ladder site $(x,y)$ ($x\in \{0,1,\hdots, N_x-1$, $y\in \{0,1 \}$). We assume periodic boundary conditions $(x+N_x, y) = (x,y)$ in the $x$-direction. From this starting point, in this section we obtain parent Hamiltonians for three variants of the cluster state on a $D=1$ dimensional chain [see Fig. \ref{fig:cluster_schematic} (right)]: the standard cluster state, a volume-law entangled state we call the rainbow cluster state (in analogy with the rainbow product of Bell pairs \cite{Lan-22a}), and another volume-law entangled state we call the antipodal cluster state (in analogy with the antipodal product of Bell pairs \cite{Chi-24a}).

\subsection{$D=1$ dimensional cluster state}

The standard cluster state on a $D=1$ dimensional chain can be obtained from the cluster state on the ladder by the mapping $(x,y) \to n = 2x + y$ of the ladder coordinates $(x,y)$ to the chain coordinate $n$ [see Fig. \ref{fig:cluster_schematic} (top right)]. This example was discussed in the main text, with the parent Hamiltonian given in Eq. \ref{eq:H_cluster}. We note that the mapping $(x,y) \to n = 2x + y$ is local, in the sense that nearby qubits on the ladder are mapped to nearby qubits on the chain.

\subsection{$D=1$ ``rainbow'' cluster state} \label{app:rainbow_cluster}

We consider a nonlocal variant of the cluster state, obtained by the nonlocal mapping $(x,y) \to n = x - 2xy + (N-1)y$ of ladder coordinates $(x,y)$ to the chain coordinate $n$. Fig. \ref{fig:cluster_schematic} (middle right) shows that this transforms the stabilizer generators of the ladder cluster state to a set of stabilizer generators given by:
\begin{eqnarray} \mathcal{S}_{\vec{\theta}}^{\rm rainbow-cluster} = \langle && \theta_0 \hat{\sigma}_{N-1}^X \hat{\sigma}_0^Z \hat{\sigma}_{\frac{N}{2}}^X, \{ \theta_n \hat{\sigma}_{N-n-1}^X \hat{\sigma}_n^Z \hat{\sigma}^X_{N-n} \}_{n \neq 0, N/2} ,  \theta_{\frac{N}{2}} \hat{\sigma}_{\frac{N}{2}-1}^X \hat{\sigma}_{\frac{N}{2}}^Z \hat{\sigma}_0^X  \rangle , \label{eq:rainbow_cluster_stab} \end{eqnarray} with periodic boundary conditions $n \equiv n + N$ on the chain.

\begin{figure}
\includegraphics[width=\columnwidth]{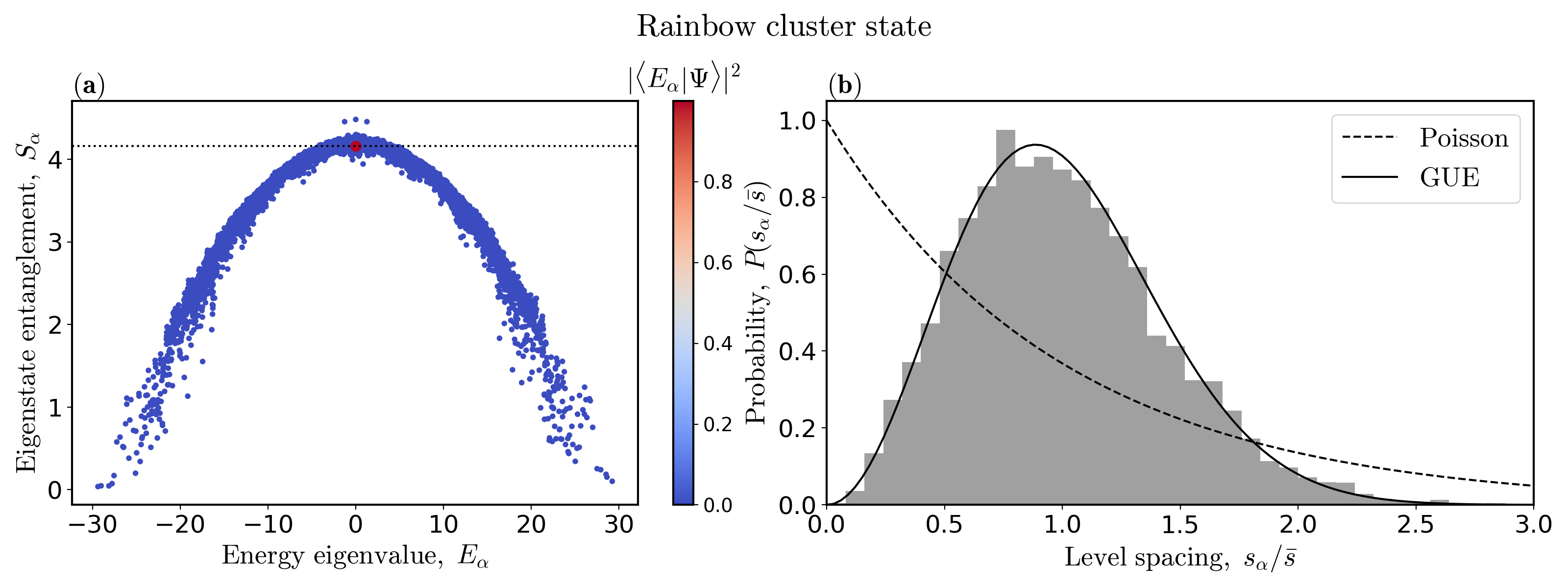}%
\caption{Numerical data for the parent Hamiltonian (given in Eq. \ref{eq:H_rainbow_cluster}) for the rainbow cluster state. (a) Eigenstate entanglement entropies, corresponding to the bipartition of the chain highlighted in gray in Fig. \ref{fig:cluster_schematic} (middle right). The rainbow cluster state (red marker) has the volume-law scaling of entanglement $S^{\rm rainbow-cluster} = \frac{1}{2}(N-2)\ln 2$ (black dotted line). (b) The distribution of level spacings verifies that the parent Hamiltonian is nonintegrable. [Parameters: $N=14$, for each $n$, $\theta_n \in \{ \pm 1 \}$ is chosen at random, and the Hamiltonian parameters are $J_n = 1.2$, $J'_n = 0.7$, $J''_n = 0.9$, $J'''_n = 0.3$, $\lambda = 0.1$, $\lambda' = -0.25$, $\eta = 0.5$, $\eta' = 0.05$, $\kappa = -0.3$.]}  
\label{fig:rainbow_cluster_example} 
\end{figure}

The stabilizer elements of $\mathcal{S}_{\vec{\theta}}^{\rm rainbow-cluster}$ are at least 3-body and are generally highly nonlocal, however, there are several $(2,2,2)$-factorizable stabilizer elements, which are listed in Table \ref{table:rainbow_cluster} along with the corresponding annihilating 2-local, 2-body operators. We also list in Table \ref{table:rainbow_cluster} some $(2,3,2)$-factorizable stabilizer elements and their corresponding $3$-local, 2-body annihilating operators. By Result \ref{result:main}, a parent Hamiltonian for the rainbow cluster state is a linear combination of all terms in the right column of Table \ref{table:rainbow_cluster}:

\begin{eqnarray} \hat{H}^{\rm rainbow-cluster} &=& \sum_{n \neq 0, N/2} J_n (\hat{\sigma}_n^Z - \theta_n \hat{\sigma}_{N-n-1}^X \hat{\sigma}_{N-n}^X) + \sum_{n \neq 0, \frac{N}{2}-1, \frac{N}{2}, N-1} J''_n (\hat{\sigma}_n^Y \hat{\sigma}_{n+1}^X - \theta_n \theta_{N-n-1} \hat{\sigma}_{N-n-1}^Y \hat{\sigma}_{N-n}^X) \nonumber \\ &+&  \sum_{n \neq 0, N/2} J'_n (\hat{\sigma}_{n-1}^X \hat{\sigma}_n^Y - \theta_n \theta_{N-n} \hat{\sigma}_{N-n-1}^X \hat{\sigma}_{N-n}^Y)  + \sum_{n \neq 0, \frac{N}{2}-1, \frac{N}{2}, N-1} J'''_n (\hat{\sigma}_n^Z \hat{\sigma}_{n+1}^Z - \theta_n \theta_{n+1} \hat{\sigma}_{N-n-2}^X \hat{\sigma}_{N-n}^X ) \nonumber \\ &+& \lambda (\hat{\sigma}_{1}^X - \theta_{N-1}\hat{\sigma}_{N-1}^Z \hat{\sigma}_{0}^X) + \lambda' (\hat{\sigma}_{\frac{N}{2}}^X - \theta_0 \hat{\sigma}_{N-1}^X \hat{\sigma}_{0}^Z) + \eta (\hat{\sigma}_{0}^X - \theta_{\frac{N}{2}}\hat{\sigma}_{\frac{N}{2}-1}^X \hat{\sigma}_{\frac{N}{2}}^Z) + \eta' (\hat{\sigma}_{\frac{N}{2}+1}^X - \theta_{\frac{N}{2}-1} \hat{\sigma}_{\frac{N}{2}-1}^Z \hat{\sigma}_{\frac{N}{2}}^X ) \nonumber \\ &+& \kappa (\hat{\sigma}_{N-1}^X \hat{\sigma}_{0}^Y - \theta_0 \theta_{\frac{N}{2}} \hat{\sigma}_{\frac{N}{2}-1}^X \hat{\sigma}_{\frac{N}{2}}^Y) , \label{eq:H_rainbow_cluster} \end{eqnarray} which is naturally interpreted as two spin chains ($0 \leq n \leq N/2 -1$ and $N/2 \leq n \leq N-1$) coupled by the terms in the last two lines of Eq. \ref{eq:H_rainbow_cluster}. In Fig. \ref{fig:rainbow_cluster_example}(a) we plot the eigenstate expectation values for this parent Hamiltonian. The rainbow cluster state (the red marker) is a volume-law entangled QMBS that is difficult to distinguish from its surrounding thermal states. However, it is nonthermal since it is a stabilizer state and has zero magic, in contrast to the surrounding thermal states.

The half-chain entanglement entropy of the rainbow state [for the bipartition highlighted in gray in Fig. \ref{fig:cluster_schematic} (middle right)] can be computed by the methods of Ref. \cite{Fat-04a}. One can see in Fig. \ref{fig:cluster_schematic} (middle right) that the product of all stabilizer generators coloured in blue and green gives the stabilizer element $\bigotimes_{n=0}^{N/2-1}\hat{\sigma}_n^Z$, which acts nontrivially only on the highlighted gray subsystem, which we call subsystem $A$. Similarly, the product of all stabilizer generators coloured in red and purple gives the stabilizer element $\bigotimes_{n=N/2}^{N-1}\hat{\sigma}_n^Z$, which acts nontrivially only on the complementary subsytem which we call $B$. An inspection of the stabilizer group shows that these are the only stabilizer elements that are entirely contained within subsystem $A$ or $B$, so that we have the subgroup $\mathcal{S}_A \times \mathcal{S}_B = \{ \hat{\mathbb{I}}^{\otimes N}, \bigotimes_{n=0}^{N/2-1}\hat{\sigma}_n^Z, \bigotimes_{n=N/2}^{N-1}\hat{\sigma}_n^Z, \bigotimes_{n=0}^{N-1}\hat{\sigma}_n^Z \} \subset \mathcal{S}^{\rm rainbow-cluster}$. Since the subgroup $\mathcal{S}_A \times \mathcal{S}_B$ contains 4 elements, the remaining subgroup $\mathcal{S}_{AB}$ contains all other stabilizer elements and therefore has the size $|\mathcal{S}_{AB}| = 2^{N-2}$. It follows, by the results of Ref. \cite{Fat-04a}, that the entanglement entropy of the rainbow cluster state is $S = \frac{1}{2} \ln |\mathcal{S}_{AB}| = \frac{1}{2} (N - 2) \ln 2$, which scales with the volume of the subsystem.

The local Hamiltonian $\hat{H}^{\rm rainbow-cluster}$ is nonintegrable as verified by the distribution of level spacings in Fig. \ref{fig:rainbow_cluster_example}(b).

\begin{table}
\begin{center} 
  \begin{tabular}{ r l l }
    \toprule
    \multicolumn{2}{c}{\bf $(2,2,2)$-factorizable} \hspace{5mm} & {\bf $2$-local, $2$-body parent}  \\ \multicolumn{2}{c}{\bf stabilizer element} \hspace{5mm} & {\bf Hamiltonian term}  \\ \midrule 
  ($ n \neq 0, N/2$)  &  &  \\  $\theta_n \hat{\sigma}_{N-n-1}^X \hat{\sigma}_n^Z \hat{\sigma}_{N-n}^X$ & $= \theta_n (\hat{\sigma}_n^Z) (\hat{\sigma}_{N-n-1}^X \hat{\sigma}_{N-n}^X )$ & $ \hat{\sigma}_n^Z - \theta_n \hat{\sigma}_{N-n-1}^X \hat{\sigma}_{N-n}^X $ \\ $(\theta_n \hat{\sigma}_{N-n-1}^X \hat{\sigma}_n^Z \hat{\sigma}_{N-n}^X)  (\theta_{N-n} \hat{\sigma}_{n-1}^X \hat{\sigma}_{N-n}^Z \hat{\sigma}_{n}^X ) $ & $= \theta_n \theta_{N-n} (\hat{\sigma}_{N-n-1}^X \hat{\sigma}_{N-n}^Y ) ( \hat{\sigma}_{n-1}^X \hat{\sigma}_n^Y ) $ & $ \hat{\sigma}_{n-1}^X \hat{\sigma}_n^Y - \theta_n \theta_{N-n} \hat{\sigma}_{N-n-1}^X \hat{\sigma}_{N-n}^Y $ \\ \midrule ($n \neq 0,\frac{N}{2}-1, \frac{N}{2}, N-1$) &  &  \\ $(\theta_n \hat{\sigma}_{N-n-1}^X \hat{\sigma}_n^Z \hat{\sigma}_{N-n}^X)  (\theta_{N-n-1} \hat{\sigma}_{n}^X \hat{\sigma}_{N-n-1}^Z \hat{\sigma}_{n+1}^X ) $ & $= \theta_n \theta_{N-n-1} (\hat{\sigma}_n^Y \hat{\sigma}_{n+1}^X ) ( \hat{\sigma}_{N-n-1}^Y \hat{\sigma}_{N-n}^X)$ & $\hat{\sigma}_n^Y \hat{\sigma}_{n+1}^X - \theta_n \theta_{N-n-1} \hat{\sigma}_{N-n-1}^Y \hat{\sigma}_{N-n}^X$ \\  \midrule $\theta_{N-1} \hat{\sigma}_{0}^X \hat{\sigma}_{N-1}^Z \hat{\sigma}_{1}^X$ & $= \theta_{N-1} (\hat{\sigma}_{1}^X) (\hat{\sigma}_{N-1}^Z \hat{\sigma}_{0}^X )$ & $\hat{\sigma}_{1}^X - \theta_{N-1}\hat{\sigma}_{N-1}^Z \hat{\sigma}_{0}^X$ \\ $\theta_0 \hat{\sigma}_0^Z \hat{\sigma}_{\frac{N}{2}}^X \hat{\sigma}_{N-1}^X$ & $= \theta_0 (\hat{\sigma}_{\frac{N}{2}}^X) (\hat{\sigma}_{N-1}^X \hat{\sigma}_{0}^Z )$ & $ \hat{\sigma}_{\frac{N}{2}}^X - \theta_0 \hat{\sigma}_{N-1}^X \hat{\sigma}_{0}^Z $ \\ $\theta_{\frac{N}{2}} \hat{\sigma}_{\frac{N}{2}-1}^X \hat{\sigma}_{\frac{N}{2}}^Z  \hat{\sigma}_{0}^X$ & $= \theta_{\frac{N}{2}} (\hat{\sigma}_{0}^X) (\hat{\sigma}_{\frac{N}{2}-1}^X \hat{\sigma}_{\frac{N}{2}}^Z )$ & $ \hat{\sigma}_{0}^X - \theta_{\frac{N}{2}}\hat{\sigma}_{\frac{N}{2}-1}^X \hat{\sigma}_{\frac{N}{2}}^Z $ \\ $\theta_{\frac{N}{2}-1} \hat{\sigma}_{\frac{N}{2}}^X \hat{\sigma}_{\frac{N}{2}-1}^Z \hat{\sigma}_{\frac{N}{2}+1}^X$ & $= \theta_{\frac{N}{2}-1} ( \hat{\sigma}_{\frac{N}{2}+1}^X ) ( \hat{\sigma}_{\frac{N}{2}-1}^Z \hat{\sigma}_{\frac{N}{2}}^X )$ & $\hat{\sigma}_{\frac{N}{2}+1}^X - \theta_{\frac{N}{2}-1} \hat{\sigma}_{\frac{N}{2}-1}^Z \hat{\sigma}_{\frac{N}{2}}^X $ \\ $(\theta_0 \hat{\sigma}_{N-1}^X \hat{\sigma}_0^Z \hat{\sigma}_{\frac{N}{2}}^X) (\theta_{\frac{N}{2}} \hat{\sigma}_0^X \hat{\sigma}_{\frac{N}{2}}^Z \hat{\sigma}_{\frac{N}{2}-1}^X)$ & $= \theta_0 \theta_{\frac{N}{2}} (\hat{\sigma}_{N-1}^X \hat{\sigma}_{0}^Y) (\hat{\sigma}_{\frac{N}{2}-1}^X \hat{\sigma}_{\frac{N}{2}}^Y )$ & $ \hat{\sigma}_{N-1}^X \hat{\sigma}_{0}^Y - \theta_0 \theta_{\frac{N}{2}} \hat{\sigma}_{\frac{N}{2}-1}^X \hat{\sigma}_{\frac{N}{2}}^Y $ \\ 
    \bottomrule
    \multicolumn{2}{c}{\bf $(2,3,2)$-factorizable} \hspace{5mm} & {\bf $3$-local, $2$-body parent}  \\ \multicolumn{2}{c}{\bf stabilizer element} \hspace{5mm} & {\bf Hamiltonian term}  \\ \midrule
    ($n \neq 0,\frac{N}{2}-1, \frac{N}{2}, N-1$) &  &  \\ $(\theta_n \hat{\sigma}_{N-n-1}^X \hat{\sigma}_n^Z \hat{\sigma}_{N-n}^X)  (\theta_{n+1} \hat{\sigma}_{N-n-2}^X \hat{\sigma}_{n+1}^Z \hat{\sigma}_{N-n-1}^X ) $ & $= \theta_n \theta_{n+1} (\hat{\sigma}_n^Z \hat{\sigma}_{n+1}^Z) ( \hat{\sigma}_{N-n-2}^X \hat{\sigma}_{N-n}^X)$ & $\hat{\sigma}_n^Z \hat{\sigma}_{n+1}^Z - \theta_n \theta_{n+1} \hat{\sigma}_{N-n-2}^X \hat{\sigma}_{N-n}^X$ \\
    \bottomrule
  \end{tabular}
\end{center}
\caption{All $(2,2,2)$-factorizable stabilizer elements [and some $(2,3,2)$-factorizable stabilizer elements] of the rainbow cluster state (with stabilizer group in Eq. \ref{eq:rainbow_cluster_stab}), and the corresponding local, 2-body operators that annihilate the rainbow cluster state.}
\label{table:rainbow_cluster}
\end{table}

\subsection{$D=1$ ``antipodal'' cluster state}

Another nonlocal variant of the $D=1$ cluster state, which we dub the antipodal cluster state, is obtained by the nonlocal mapping $(x,y) \to n = x + \frac{N}{2}y$ of ladder coordinates $(x,y)$ to the chain coordinate $n$. The antipodal cluster state is defined by its stabilizer generators (assuming $N$ even and periodic boundary conditions $n \equiv n + N$):
\begin{eqnarray} \mathcal{S}_{\vec{\theta}}^{\rm antipodal-cluster} = \langle && \theta_0 \hat{\sigma}_{N-1}^X \hat{\sigma}_0^Z \hat{\sigma}_{\frac{N}{2}}^X, \nonumber \\ && \{ \theta_n \hat{\sigma}_{n+\frac{N}{2}-1}^X \hat{\sigma}_n^Z \hat{\sigma}^X_{n+\frac{N}{2}} \}_{1 \leq n \leq \frac{N}{2}-1} , \nonumber \\ && \{ \theta_n \hat{\sigma}_{n-\frac{N}{2}}^X \hat{\sigma}_n^Z \hat{\sigma}^X_{n-\frac{N}{2}+1} \}_{\frac{N}{2} \leq n \leq N-2} , \nonumber \\ && \theta_{N-1} \hat{\sigma}_{\frac{N}{2}-1}^X \hat{\sigma}_{N-1}^Z \hat{\sigma}_0^X  \rangle . \label{eq:antipodal_cluster_stab} \end{eqnarray}
which are illustrated in Fig. \ref{fig:cluster_schematic} (bottom right).
 
\begin{figure}
\includegraphics[width=\columnwidth]{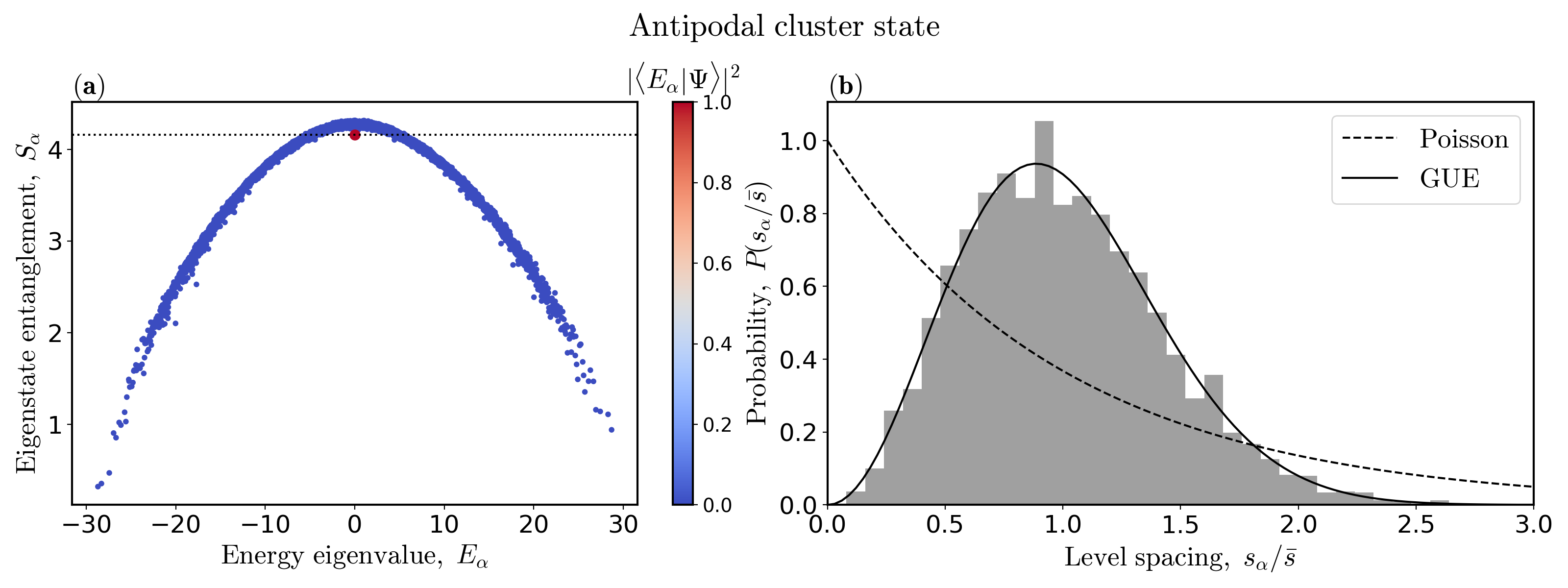}%
\caption{Numerical data for the parent Hamiltonian $\hat{H}^{\rm antipodal-cluster}$ (Eq. \ref{eq:H_antipodal_cluster}) of the antipodal cluster state. (a) Eigenstate entanglement entropies, corresponding to the bipartition of the chain highlighed in gray in Fig. \ref{fig:cluster_schematic} (bottom right). The antipodal cluster state (red marker) has the volume-law scaling of entanglement $S^{\rm rainbow-cluster} = \frac{1}{2}(N-2)\ln 2$ (black dotted line). (b) The distribution of level spacings verifies that the parent Hamiltonian is nonintegrable. [Parameters: $N=14$, $\vec{\theta} = (1,1,\hdots,1)$ and $\lambda = 1$, $\lambda' = 1.3$, $J = 0.5$, $J' = 1.7$, $\omega = 1.2$, $\omega' = 0.9$, $\Omega = 1.6$.]} 
\label{fig:example_antipodal_cluster} 
\end{figure}

Each stabilizer element of $\mathcal{S}_{\vec{\theta}}^{\rm antipodal-cluster}$ is at least 3-body and $\ell$-local, with $\ell = N/2$. Despite this, there are several $(2,2,2)$-factorizable stabilizer elements, which are listed in Table \ref{table:antipodal_cluster} along with the corresponding annihilating 2-local, 2-body operators. By Result \ref{result:main}, the 2-local, 2-body parent Hamiltonian for the antipodal cluster state is a linear combination of all terms in the right column of Table \ref{table:antipodal_cluster}:
\begin{eqnarray} \hat{H}^{\rm antipodal-cluster} &=& \sum_{n=1}^{N/2 - 1} \lambda_n (\hat{\sigma}_n^Z - \theta_n \hat{\sigma}_{n+\frac{N}{2}-1}^X \hat{\sigma}_{n+\frac{N}{2}}^X) + \sum_{n=1}^{N/2 - 1} \lambda'_n (\hat{\sigma}_{n+\frac{N}{2}-1}^Z - \theta_{n+\frac{N}{2}-1}\hat{\sigma}_{n-1}^X \hat{\sigma}_{n}^X) \nonumber \\ &+& \sum_{n=1}^{N/2 - 1} J_n (\hat{\sigma}_{n-1}^X \hat{\sigma}_{n}^Y - \theta_n \theta_{n+\frac{N}{2}-1} \hat{\sigma}_{n+\frac{N}{2}-1}^Y \hat{\sigma}_{n+\frac{N}{2}}^X ) + \sum_{n=1}^{N/2 - 2} J'_n (\hat{\sigma}_n^Y \hat{\sigma}_{n+1}^X - \theta_n \theta_{n+\frac{N}{2}} \hat{\sigma}_{n+\frac{N}{2}-1}^X \hat{\sigma}_{n+\frac{N}{2}}^Y) \nonumber \\ &+& \omega (\hat{\sigma}_{\frac{N}{2}}^X - \theta_0 \hat{\sigma}_{N-1}^X \hat{\sigma}_{0}^Z) + \omega' (\hat{\sigma}_{\frac{N}{2}-1}^X - \theta_{N-1} \hat{\sigma}_{N-1}^Z \hat{\sigma}_{0}^X) + \Omega (\hat{\sigma}_{N-1}^Y \hat{\sigma}_{0}^Y - \theta_0 \theta_{N-1} \hat{\sigma}_{\frac{N}{2}-1}^X \hat{\sigma}_{\frac{N}{2}}^X) . \end{eqnarray}
Assuming $\theta_n = 1$ for all $n$, and that the Hamiltonian parameters $\lambda_n = \lambda$, $\lambda'_n = \lambda'$, $J_n = J$, $J'_n = J'$ are all independent of $n$, gives the somewhat simplified parent Hamiltonian: \begin{equation} \hat{H}^{\rm antipodal-cluster} = \hat{H}^{(L)} + \hat{H}^{(R)} + \hat{V}^{(LR)} + \hat{V}^{(RL)} , \label{eq:H_antipodal_cluster} \end{equation} where: \begin{eqnarray} \hat{H}^{(L)} &=& \lambda \sum_{n=1}^{\frac{N}{2}-1} \hat{\sigma}_n^Z + \lambda' \sum_{n=1}^{\frac{N}{2}-1} \hat{\sigma}_{n-1}^X \hat{\sigma}_n^X  + J \sum_{n=1}^{\frac{N}{2}-1} \hat{\sigma}_{n-1}^X \hat{\sigma}_{n}^Y + J' \sum_{n=1}^{\frac{N}{2}-2} \hat{\sigma}_{n}^Y \hat{\sigma}_{n+1}^X  \nonumber \\ \hat{H}^{(R)} &=& \lambda' \sum_{n=\frac{N}{2}}^{N-2} \hat{\sigma}_n^Z - \lambda \sum_{n=\frac{N}{2}}^{N-2} \hat{\sigma}_{n}^X \hat{\sigma}_{n+1}^X - J \sum_{n=\frac{N}{2}}^{N-2} \hat{\sigma}_n^Y \hat{\sigma}_{n+1}^X  - J' \sum_{n=\frac{N}{2}}^{N-3} \hat{\sigma}_{n}^X \hat{\sigma}_{n+1}^Y \nonumber \\ \hat{V}^{(LR)} &=& \Omega \hat{\sigma}_{N-1}^Y \hat{\sigma}_0^Y - \omega \hat{\sigma}_{N-1}^X \hat{\sigma}_0^Z - \omega' \hat{\sigma}_{N-1}^Z \hat{\sigma}_0^X \\  \hat{V}^{(RL)} &=& \omega \hat{\sigma}_{\frac{N}{2}}^X + \omega' \hat{\sigma}_{\frac{N}{2}-1}^X - \Omega \hat{\sigma}_{\frac{N}{2}-1}^X \hat{\sigma}_{\frac{N}{2}}^X \end{eqnarray} This is the Hamiltonian for two qubit chains (with Hamiltonians $\hat{H}^{(L)}$ for the qubits $0 \leq n \leq N/2$ and $\hat{H}^{(R)}$ for the qubits $N/2+1 \leq n \leq N-1$) coupled at both ends by the interactions $\hat{V}^{(LR)}$ and $\hat{V}^{(RL)}$. Fig. \ref{fig:example_antipodal_cluster} shows the half-chain entanglement entropy of each eigenstate of this parent Hamiltonian. The antipodal cluster state (the red marker) is a volume-law entangled QMBS, like its surrounding thermal states. However, it is nonthermal since it is a stabilizer state and has zero magic, in contrast to the surrounding thermal states. Although the Hamiltonian $\hat{H}^{(L)} + \hat{H}^{(R)}$ for the two noninteracting chains is integrable (it can be mapped to free fermions) the interaction $\hat{V}^{(LR)} + \hat{V}^{(RL)}$ between the chains break the integrability, as verified by the distribution of level spacings in Fig. \ref{fig:example_antipodal_cluster}(b).

Similar to the argument in Sec. \ref{app:rainbow_cluster} for the rainbow cluster state, the half-chain entanglement entropy of the antipodal cluster state [for the bipartition highlighted in gray in Fig. \ref{fig:cluster_schematic} (bottom right)] can be computed by the methods of Ref. \cite{Fat-04a}. One can see in Fig. \ref{fig:cluster_schematic} (bottom right) that the product of all stabilizer generators coloured in blue and green gives the stabilizer element $\bigotimes_{n=0}^{N/2-1}\hat{\sigma}_n^Z$, which acts nontrivially only on the highlighted gray subsystem, which we call subsystem $A$. Similarly, the product of all stabilizer generators coloured in red and purple gives the stabilizer element $\bigotimes_{n=N/2}^{N-1}\hat{\sigma}_n^Z$, which acts nontrivially only on the complementary subsytem which we call $B$. An inspection of the stabilizer group shows that these are the only stabilizer elements that act nontrivially only on subsystem $A$ or $B$, so that we have the subgroup $\mathcal{S}_A \times \mathcal{S}_B = \{ \hat{\mathbb{I}}^{\otimes N}, \bigotimes_{n=0}^{N/2-1}\hat{\sigma}_n^Z, \bigotimes_{n=N/2}^{N-1}\hat{\sigma}_n^Z, \bigotimes_{n=0}^{N-1}\hat{\sigma}_n^Z \} \subset \mathcal{S}^{\rm antipodal-cluster}$. Since the subgroup $\mathcal{S}_A \times \mathcal{S}_B$ contains 4 elements, the remaining subgroup $\mathcal{S}_{AB}$ contains all other stabilizer elements and therefore has the size $|\mathcal{S}_{AB}| = 2^{N-2}$. It follows, by the results of Ref. \cite{Fat-04a}, that the entanglement entropy of the antipodal cluster state is $S = \frac{1}{2} \ln |\mathcal{S}_{AB}| = \frac{1}{2} (N - 2) \ln 2$, which scales with the volume of the subsystem.

\begin{table}
\begin{center} 
  \begin{tabular}{ r l l }
    \toprule
    \multicolumn{2}{c}{\bf $(2,2,2)$-factorizable stabilizer element} \hspace{5mm} & {\bf $2$-local, $2$-body parent}  \\ \multicolumn{2}{c}{ } \hspace{5mm} & {\bf Hamiltonian term}  \\ \midrule 
  ($1 \leq n \leq \frac{N}{2}-1$)  &  &  \\  $\theta_n \hat{\sigma}_n^Z \hat{\sigma}_{n+\frac{N}{2}-1}^X \hat{\sigma}_{n+\frac{N}{2}}^X$ & $= \theta_n (\hat{\sigma}_n^Z) (\hat{\sigma}_{n+\frac{N}{2}-1}^X \hat{\sigma}_{n+\frac{N}{2}}^X )$ & $ \hat{\sigma}_n^Z - \theta_n \hat{\sigma}_{n+\frac{N}{2}-1}^X \hat{\sigma}_{n+\frac{N}{2}}^X $ \\ $\theta_{n+\frac{N}{2}-1} \hat{\sigma}_{n+\frac{N}{2}-1}^Z \hat{\sigma}_{n-1}^X \hat{\sigma}_{n}^X$ & $= \theta_{n+\frac{N}{2}-1} (\hat{\sigma}_{n+\frac{N}{2}-1}^Z ) (\hat{\sigma}_{n-1}^X \hat{\sigma}_{n}^X)$ & $\hat{\sigma}_{n+\frac{N}{2}-1}^Z - \theta_{n+\frac{N}{2}-1}\hat{\sigma}_{n-1}^X \hat{\sigma}_{n}^X$ \\ $(\theta_n \hat{\sigma}_n^Z \hat{\sigma}_{n+\frac{N}{2}-1}^X \hat{\sigma}_{n+\frac{N}{2}}^X)  (\theta_{n+\frac{N}{2}-1} \hat{\sigma}_{n+\frac{N}{2}-1}^Z \hat{\sigma}_{n-1}^X \hat{\sigma}_{n}^X ) $ & $= \theta_n \theta_{n+\frac{N}{2}-1} (\hat{\sigma}_{n-1}^X \hat{\sigma}_{n}^Y) (\hat{\sigma}_{n+\frac{N}{2}-1}^Y \hat{\sigma}_{n+\frac{N}{2}}^X ) $ & $\hat{\sigma}_{n-1}^X \hat{\sigma}_{n}^Y - \theta_n \theta_{n+\frac{N}{2}-1} \hat{\sigma}_{n+\frac{N}{2}-1}^Y \hat{\sigma}_{n+\frac{N}{2}}^X $ \\ \midrule ($1 \leq n \leq \frac{N}{2}-2$) &  &  \\ $(\theta_n \hat{\sigma}_n^Z \hat{\sigma}_{n+\frac{N}{2}-1}^X \hat{\sigma}_{n+\frac{N}{2}}^X)(\theta_{n+\frac{N}{2}} \hat{\sigma}_{n+\frac{N}{2}}^Z \hat{\sigma}_{n}^X \hat{\sigma}_{n+1}^X)$ & $= \theta_n \theta_{n+\frac{N}{2}} (\hat{\sigma}_n^Y \hat{\sigma}_{n+1}^X) (\hat{\sigma}_{n+\frac{N}{2}-1}^X \hat{\sigma}_{n+\frac{N}{2}}^Y )$ & $\hat{\sigma}_n^Y \hat{\sigma}_{n+1}^X - \theta_n \theta_{n+\frac{N}{2}} \hat{\sigma}_{n+\frac{N}{2}-1}^X \hat{\sigma}_{n+\frac{N}{2}}^Y$ \\  \midrule $\theta_0 \hat{\sigma}_0^Z \hat{\sigma}_{\frac{N}{2}}^X \hat{\sigma}_{N-1}^X$ & $= \theta_0 (\hat{\sigma}_{\frac{N}{2}}^X) (\hat{\sigma}_{N-1}^X \hat{\sigma}_{0}^Z )$ & $ \hat{\sigma}_{\frac{N}{2}}^X - \theta_0 \hat{\sigma}_{N-1}^X \hat{\sigma}_{0}^Z $ \\ $\theta_{N-1} \hat{\sigma}_0^X \hat{\sigma}_{\frac{N}{2}-1}^X \hat{\sigma}_{N-1}^Z$ & $= \theta_{N-1} (\hat{\sigma}_{\frac{N}{2}-1}^X) (\hat{\sigma}_{N-1}^Z \hat{\sigma}_{0}^X )$ & $ \hat{\sigma}_{\frac{N}{2}-1}^X - \theta_{N-1} \hat{\sigma}_{N-1}^Z \hat{\sigma}_{0}^X $ \\ $(\theta_0 \hat{\sigma}_0^Z \hat{\sigma}_{\frac{N}{2}}^X \hat{\sigma}_{N-1}^X) (\theta_{N-1} \hat{\sigma}_0^X \hat{\sigma}_{\frac{N}{2}-1}^X \hat{\sigma}_{N-1}^Z)$ & $= \theta_0 \theta_{N-1} (\hat{\sigma}_{N-1}^Y \hat{\sigma}_{0}^Y) (\hat{\sigma}_{\frac{N}{2}-1}^X \hat{\sigma}_{\frac{N}{2}}^X )$ & $ \hat{\sigma}_{N-1}^Y \hat{\sigma}_{0}^Y - \theta_0 \theta_{N-1} \hat{\sigma}_{\frac{N}{2}-1}^X \hat{\sigma}_{\frac{N}{2}}^X $ \\
  \bottomrule
  \end{tabular}
\end{center}
\caption{All $(2,2,2)$-factorizable stabilizer elements of the antipodal cluster state (with stabilizer group in Eq. \ref{eq:antipodal_cluster_stab}), and the corresponding 2-local, 2-body operators that annihilate the antipodal cluster state.}
\label{table:antipodal_cluster}
\end{table}

\section{Stabilizer eigenstate of the PXP Hamiltonian} \label{app:PXP}

The PXP Hamiltonian on a chain of $N$ qubits is: \begin{equation} \hat{H}^{\rm PXP} = \sum_{n=0}^{N-1} \hat{\mathcal{P}}_{n-1} \hat{\sigma}_n^X \hat{\mathcal{P}}_{n+1} , \end{equation} where $\hat{\mathcal{P}}_n = \frac{1}{2} (\hat{\mathbb{I}} - \hat{\sigma}_n^Z)$. Substituting the expression for $\hat{\mathcal{P}}_n$ into $\hat{H}^{\rm PXP}$ allows us to write the PXP Hamiltonian as a linear combination of Pauli strings: \begin{equation} \hat{H}^{\rm PXP} = \frac{1}{4} \sum_{n=0}^{N-1} ( \hat{\sigma}_n^X - \hat{\sigma}_n^X \hat{\sigma}_{n+1}^Z - \hat{\sigma}_{n-1}^Z \hat{\sigma}_n^X + \hat{\sigma}_{n-1}^Z \hat{\sigma}_n^X \hat{\sigma}_{n+1}^Z ) . \end{equation} It is known that the antipodal product of Bell pairs is an eigenstate of the PXP Hamiltonian \cite{Iva-25a, Chi-24a}, which we now show using our framework.

The antipodal product of Bell pairs (Eq. \ref{eq:psi_antipodal_Bell}) has the stabilizer group given in Eq. \ref{eq:antipodal_Bell_stab}, which we repeat here for the reader's convenience:
\begin{equation} \mathcal{S}^{\rm antipodal-BP} = \langle \{ \hat{Q}_n , \hat{Q}_{n+N/2}  \}_{n=0}^{N/2 - 1} \rangle \quad {\rm where} \quad \hat{Q}_n \equiv \theta_{n} \hat{\sigma}_{n}^X \hat{\sigma}_{n+N/2}^X \quad {\rm and} \quad \hat{Q}_{n+N/2} \equiv \theta_{n+N/2} \hat{\sigma}_{n}^Z \hat{\sigma}_{n+N/2}^Z . \end{equation} Since $\hat{Q}_n = \theta_{n} \hat{\sigma}_{n}^X \hat{\sigma}_{n+N/2}^X$ is a $(2,1,1)$-factorizable stabilizer element, the antipodal product of Bell pairs is annihilated by the 1-local, 1-body operators: \begin{equation} \hat{\sigma}_{n}^X - \theta_{n} \hat{\sigma}_{n+N/2}^X , \quad n \in \{ 0,1,\hdots, N/2-1 \} . \label{eq:PXP_1body} \end{equation} Also, since the products of generators $\hat{Q}_n \hat{Q}_{n+N/2+1} = \theta_n \theta_{n+N/2+1} (\hat{\sigma}_{n}^X \hat{\sigma}_{n+1}^Z) (\hat{\sigma}_{n+N/2}^X \hat{\sigma}_{n+N/2+1}^Z)$ and $\hat{Q}_{n+1} \hat{Q}_{n+N/2} = \theta_{n+1} \theta_{n+N/2} (\hat{\sigma}_{n}^Z \hat{\sigma}_{n+1}^X) (\hat{\sigma}_{n+N/2}^Z \hat{\sigma}_{n+N/2+1}^X)$ are $(2,2,2)$-factorizable stabilizer elements, the antipodal product of Bell pairs is annihilated by the 2-local, 2-body operators: \begin{equation} \hat{\sigma}_{n}^X \hat{\sigma}_{n+1}^Z - \theta_n \theta_{n+N/2+1} \hat{\sigma}_{n+N/2}^X \hat{\sigma}_{n+N/2+1}^Z, \quad \hat{\sigma}_{n}^Z \hat{\sigma}_{n+1}^X - \theta_{n+1} \theta_{n+N/2} \hat{\sigma}_{n+N/2}^Z \hat{\sigma}_{n+N/2+1}^X , \quad n \in \{ 0,1,\hdots, N/2-1 \} . \label{eq:PXP_2body} \end{equation} Finally, the product of generators \begin{eqnarray} \hat{Q}_{n+N/2-1} \hat{Q}_n \hat{Q}_{n+N/2+1} &=& (\theta_{n+N/2-1} \hat{\sigma}_{n-1}^Z \hat{\sigma}_{n+N/2-1}^Z) (\theta_{n} \hat{\sigma}_{n}^X \hat{\sigma}_{n+N/2}^X) (\theta_{n+N/2+1} \hat{\sigma}_{n+1}^Z \hat{\sigma}_{n+N/2+1}^Z) \nonumber \\ &=& \theta_{n+N/2-1} \theta_{n} \theta_{n+N/2+1} (\hat{\sigma}_{n-1}^Z \hat{\sigma}_{n}^X \hat{\sigma}_{n+1}^Z) (\hat{\sigma}_{n+N/2-1}^Z \hat{\sigma}_{n+N/2}^X \hat{\sigma}_{n+N/2+1}^Z) \nonumber \end{eqnarray} is $(2,3,3)$-local, so that the antipodal product of Bell pairs is annihilated by the 3-local, 3-body operators: \begin{equation}  \hat{\sigma}_{n-1}^Z \hat{\sigma}_{n}^X \hat{\sigma}_{n+1}^Z - \theta_{n+N/2-1} \theta_{n} \theta_{n+N/2+1} \hat{\sigma}_{n+N/2-1}^Z \hat{\sigma}_{n+N/2}^X \hat{\sigma}_{n+N/2+1}^Z , \quad n \in \{ 0,1,\hdots, N/2-1 \} . \label{eq:PXP_3body} \end{equation}

A linear combination of the terms in Eqs. \ref{eq:PXP_1body}, \ref{eq:PXP_2body} and \ref{eq:PXP_3body} gives a parent Hamiltonian: \begin{eqnarray} \hat{H} = \sum_{n=0}^{N/2 - 1} [&J_n^X& (\hat{\sigma}_{n}^X - \theta_{n} \hat{\sigma}_{n+N/2}^X) \nonumber \\ + &J_n^{XZ}& (\hat{\sigma}_{n}^X \hat{\sigma}_{n+1}^Z - \theta_n \theta_{n+N/2+1} \hat{\sigma}_{n+N/2}^X \hat{\sigma}_{n+N/2+1}^Z) \nonumber \\ + &J_n^{ZX}& (\hat{\sigma}_{n}^Z \hat{\sigma}_{n+1}^X - \theta_{n+1} \theta_{n+N/2} \hat{\sigma}_{n+N/2}^Z \hat{\sigma}_{n+N/2+1}^X) \nonumber \\ +  &J_n^{ZXZ}& (\hat{\sigma}_{n-1}^Z \hat{\sigma}_{n}^X \hat{\sigma}_{n+1}^Z - \theta_{n+N/2-1} \theta_{n} \theta_{n+N/2+1} \hat{\sigma}_{n+N/2-1}^Z \hat{\sigma}_{n+N/2}^X \hat{\sigma}_{n+N/2+1}^Z) ] . \end{eqnarray} for the antipodal product of Bell pairs, where $J_n^X$, $J_n^{XZ}$, $J_n^{ZX}$ and $J_n^{ZXZ}$ are arbitrary real parameters. If we chose the stabilizer phases $\theta_n = -1$ and $\theta_{n+N/2}=1$ for $n \in \{0,1,\hdots,N/2-1\}$, and also the Hamiltonian parameters $J_n^X = J_n^{ZXZ} = - J_n^{XZ} = - J_n^{ZX} = 1/4$, then the Hamiltonian simplifies to: \begin{equation} \hat{H} = \frac{1}{4} \sum_{n=0}^{N-1} (\hat{\sigma}_n^X - \hat{\sigma}_n^X \hat{\sigma}_{n+1}^Z - \hat{\sigma}_{n-1}^Z \hat{\sigma}_n^X + \hat{\sigma}_{n-1}^Z \hat{\sigma}_n^X \hat{\sigma}_{n+1}^Z) = \hat{H}^{\rm PXP} , \end{equation} which is the PXP Hamiltonian. Using Eq. \ref{eq:psi_antipodal_Bell}, the antipodal product of Bell pairs that is annihilated by $\hat{H}^{\rm PXP}$ is explicitly written as: \begin{eqnarray} \ket{\Psi^{\rm antipodal-BP}} &=& \bigotimes_{n=0}^{N/2-1} \ket{\psi (-1,1)}_{n,n+N/2} \nonumber \\ &=& \bigotimes_{n=0}^{N/2 - 1} \left[ \frac{1}{\sqrt{2}} (\ket{1}_n \ket{1}_{n+N/2} - \ket{-1}_n \ket{-1}_{n+N/2} ) \right] . \end{eqnarray}
This (volume-law entangled) exact QMBS of the PXP chain was already known from Refs. \cite{Iva-25a, Chi-24a}.

\end{document}